\documentclass{aa}
\usepackage[varg]{txfonts}

\usepackage{graphicx}
\usepackage{natbib}
\bibpunct{(}{)}{;}{a}{}{,} 
\usepackage{color}
\usepackage{float}
\usepackage{hyperref}
\usepackage{longtable}
\usepackage{rotating}
\usepackage{lscape}
\usepackage{soul}
\usepackage{subcaption}

\usepackage{xcolor}

\newcommand{\lya}{\mbox{Ly$\alpha$}}

\newcommand{\flcgs}{\mbox{erg s$^{-1}$ cm$^{-2}$}}
\newcommand{\sbl}{\mbox{erg s$^{-1}$ cm$^{-2}$ arcsec$^{-2}$}}

\newcommand{\hii}{\mbox{H\,{\scshape ii}}}
\newcommand{\hi}{\mbox{H\,{\scshape i}}}

\newcommand{\udft}{\textsf{udf-10}}
\newcommand{\mosaic}{\textsf{mosaic}}

\begin{document}

\title{The MUSE Hubble Ultra Deep Field Survey}
\subtitle{XIII. Spatially resolved spectral properties of Lyman $\rm \alpha$ haloes around star-forming galaxies at $z>3$}

\author{Floriane Leclercq\inst{\ref{inst1},\ref{inst2}\thanks{e-mail: floriane.leclercq@unige.ch}}
\and Roland Bacon\inst{\ref{inst1}}
\and Anne Verhamme\inst{\ref{inst2}}
\and Thibault Garel\inst{\ref{inst1},\ref{inst2}}
\and Jérémy Blaizot\inst{\ref{inst2}}
\and Jarle Brinchmann\inst{\ref{inst3},\ref{inst4}}
\and Sebastiano Cantalupo\inst{\ref{inst5}}
\and Adélaïde Claeyssens\inst{\ref{inst1}}
\and Simon Conseil\inst{\ref{inst1}}
\and Thierry Contini\inst{\ref{inst6}}
\and Takuya Hashimoto\inst{\ref{inst7}}
\and Edmund Christian Herenz\inst{\ref{inst8}}
\and Haruka Kusakabe\inst{\ref{inst2}}
\and Raffaella Anna Marino\inst{\ref{inst5}}
\and Michael Maseda\inst{\ref{inst4}}
\and Jorryt Matthee\inst{\ref{inst5}}
\and Peter Mitchell\inst{\ref{inst4}}
\and Gabriele Pezzuli\inst{\ref{inst5}}
\and Johan Richard\inst{\ref{inst1}}
\and Kasper Borello Schmidt\inst{\ref{inst9}}
\and Lutz Wisotzki\inst{\ref{inst9}}
}

\institute{Univ Lyon, Univ Lyon1, Ens de Lyon, CNRS, Centre de Recherche Astrophysique de Lyon UMR5574, F-69230, Saint-Genis-Laval, France\label{inst1}
\and
Observatoire de Genève, Universite de Genève, 51 Ch. des Maillettes, 1290 Versoix, Switzerland \label{inst2}
\and
Instituto de Astrof{\'i}sica e Ci{\^e}ncias do Espaço, Universidade do Porto, CAUP, Rua das Estrelas, PT4150-762 Porto, Portugal \label{inst3}
\and
Leiden Observatory, Leiden University, P.O. Box 9513, 2300 RA Leiden, The Netherlands\label{inst4}
\and
Department of Physics, ETH Z\"urich, Wolfgang-Pauli-Strasse 27, 8093 Z\"urich, Switzerland \label{inst5}
\and
Institut de Recherche en Astrophysique et Planétologie (IRAP), Université de Toulouse, CNRS, UPS, 31400 Toulouse, France \label{inst6}
\and
Tomonaga Center for the History of the Universe (TCHoU), Faculty of Pure and Applied Sciences, University of Tsukuba, Tsukuba, Ibaraki 305-8571, JAPAN  \label{inst7}
\and 
ESO Vitacura, Alonso de Córdova 3107,Vitacura, Casilla 19001, Santiago de Chile, Chile \label{inst8}
\and
Leibniz-Institut fur Astrophysik Potsdam (AIP), An der Sternwarte 16, 14482 Potsdam, Germany\label{inst9}
}

\date{Accepted 6 February 2020}


\abstract
{
We present spatially resolved maps of six individually-detected Lyman $\alpha$ haloes (LAHs) as well as a first statistical analysis of the Lyman $\alpha$ ($\lya$) spectral signature in the circum-galactic medium of high-redshift star-forming galaxies ($-17.5$ > $M_{\rm UV}$ > $-21.5$) using the Multi-Unit Spectroscopic Explorer (MUSE). Our resolved spectroscopic analysis of the LAHs reveals significant intrahalo variations of the $\lya$ line profile. 
Using a three-dimensional two-component model for the $\lya$ emission, we measured the full width at half maximum (\mbox{FWHM}), the peak velocity shift, and the asymmetry of the $\lya$ line in the core and in the halo of 19 galaxies.
We find that the $\lya$ line shape is statistically different in the halo compared to the core (in terms of width, peak wavelength, and asymmetry) for $\approx$40\% of our galaxies.
Similarly to object-by-object based studies and a recent resolved study using lensing, we find a correlation between the peak velocity shift and the width of the $\lya$ line both at the interstellar and circum-galactic scales. This trend has been predicted by radiative transfer simulations of galactic winds as a result of resonant scattering in outflows.
While there is a lack of correlation between the spectral properties and the spatial scale lengths of our LAHs, we find a correlation between the width of the line in the LAH and the halo flux fraction.  
Interestingly, UV bright galaxies ($M_{\rm UV}$ < $-20$) show broader, more redshifted, and less asymmetric $\lya$ lines in their haloes. The most significant correlation found is for the \mbox{FWHM} of the line and the UV continuum slope of the galaxy, suggesting that the redder galaxies have broader $\lya$ lines.
The generally broad and red line shapes found in the halo component suggest that the $\lya$ haloes are powered either by scattering processes through an outflowing medium, fluorescent emission from outflowing cold clumps of gas, or a mix of both.
Considering the large diversity of the $\lya$ line profiles observed in our sample and the lack of strong correlation, the interpretation of our results is still broadly open and underlines the need for realistic spatially resolved models of the LAHs.
}

\maketitle

\section{Introduction}
\label{sec:1}

Lyman $\alpha$ $\lambda$1215.67 ($\lya$) emission is the brightest recombination line of the hydrogen atom. As such, a number of galaxies at various redshifts have been discovered through the detection of the $\lya$ line (e.g., \citealt{CH98,MR02,Ou03,Ou08,Ou10}). $\lya$ emission has therefore become a prime tool for the exploration of the Universe and, particularly, in the study of the distant and faintest star-forming galaxies which represent the bulk of the galaxy population in the early Universe.
 
While historically it was seen simply as a powerful tool for detecting high redshift objects, it is now possible to use $\lya$ emission as a tracer of the interstellar medium (ISM) of galaxies as well as their gaseous envelopes, known as the circum-galactic medium (CGM). 
It also traces the gas exchanges between a galaxy and the environment which primarily drives its evolution. 
The cold circum-galactic gas is observed in emission in the form of a $\lya$ halo (LAH) which reflects the properties of the medium in terms of \hi\ gas kinematics, column density, and opacity.

Detecting diffuse LAHs around high-redshift star-forming galaxies is very challenging because of sensitivity limitations. The first tentative detections of extended $\lya$ emission using narrowband (NB) imaging were reported twenty years ago by \cite{MW98}. In order to counteract those limitations, a number of authors adopted stacking methods (e.g., \citealt{F01, H04,Ma12,Mo14,X17}) and provided statistical evidence for the ubiquity of extended $\lya$ emission around classical $\lya$ emitters (LAE). Yet these statistical methods have been criticized and contradictory results reported (e.g., \citealt{B10,F13}) casting doubts on the existence of the LAHs. 
While extended $\lya$ emission has then been observed around individual massive galaxies and quasars (e.g., \citealt{S00,M04,MDB08,M11,S11,C12}), the detection of LAHs around low-mass star-forming galaxies was difficult and required expensive procedure, such as the acquisition of an ultra-deep long-slit observation \citep{R08} or the use of the magnification power of gravitational lensing \citep{S07}.   

When it was installed at the Very Large Telescope (VLT/ESO) three years ago, the Multi-Unit Spectroscopic Explorer (MUSE, \citealt{Ba10}), thanks to its unrivaled sensitivity, revolutionized the study of the cold CGM of star-forming galaxies by allowing the detection of LAHs around 21 individual low-mass and distant galaxies (\citealt{W16}; hereafter W16) in the MUSE \textit{Hubble} Deep Field South \citep{B15}. One year later, we (\citealt{L17}; hereafter L17) extended the W16 LAE sample by a factor of ten using the MUSE \textit{Hubble} Ultra Deep Field (UDF, \citealt{B17}) and reported the detection of LAHs around 80\% of our tested sample of LAEs. This result showed the ubiquity of the $\lya$ haloes around star-forming galaxies at redshift 3 < $z$ < 6, suggesting that there are significant hydrogen reservoirs in their CGM. 
In the L17 study, we undertook a statistical study of the spatial properties of the detected LAHs. We found that the $\lya$ emission has a median exponential scale length of 4.5 physical kpc and is, on average, ten times more extended than the UV continuum of the host galaxy. We also performed a detailed investigation of the origin of the LAHs and concluded that we were not able to disentangle the possible mechanisms: scattering from star-forming regions, fluorescence, cooling radiation from cold gas accretion, and emission from satellite galaxies \citep{G96,K96,H00,H01,C05,D06,K10,Bar10,L15}. 

Another way to get information about the CGM of distant galaxies is to study the spectral properties of the diffuse $\lya$ emission. Following the pioneering work of \cite{S07} and using the MUSE instrument coupled to the magnification power of lensing, such investigations have recently been carried out for a few $z>3$ galaxies \citep{Sm17,Ven17}. Those highly magnified systems allow a well-resolved analysis of the $\lya$ spectral properties. Unfortunately, such systems are very rare at high redshift and rely on lens modeling. 
Until today and aside from very extended $\lya$ blobs (e.g., \citealt{Ven17}), significant spatial variations in the $\lya$ line profiles have been reported within only three lensed LAHs at $z >$ 3.5 (\citealt{P16}, \citealt{Sm17}, \citealt{C19}; hereafter C19) and one un-lensed extended LAE at $z\simeq$ 2 \citep{E18}. The resolved study of \cite{E18} reports the detection of spatial variation in the peak ratio and peak separation of a double-peaked $\lya$ emitter. While the red peak is dominant at the center of the galaxy, the peak ratio becomes close to unity at the outskirts of the halo. The authors indicated that such observations agree with the presence of outflows in the system. They also measured variations of $\approx$300 km s$^{-1}$ in the $\lya$ peak separation which they interpreted as variations in the medium properties in terms of column density, covering fraction or velocity. The C19 analysis presents detailed maps of $\lya$ line properties (velocity peak shift and width) at sub-kpc scales for two lensed LAEs. For the two objects, significant spatial variations in the line shape were found, as well as a global trend for the $\lya$ line to be redder and wider at large radii. 

Following those recent studies, here we go further into the analysis of the diffuse $\lya$ emission surrounding non-lensed galaxies by looking at the spatially resolved properties of the $\lya$ haloes detected in our previous study L17. 
A significant improvement in the data reduction of the MUSE UDF data \citep{B19} allows for a detailed analysis of the brightest LAHs. 
By taking advantage of the three-dimensional (3D) information provided by the MUSE data cubes, we first look at spectral variations by creating spatially resolved maps of the $\lya$ surface brightness and line properties. Then, we adopt an approach similar to L17 and describe the $\lya$ distribution with two components, but which now also consider the spectral dimension. This 3D parametric method allows us to disentangle the contribution of the central and diffuse components to the total $\lya$ spectral signature. The connection between the spectral and spatial properties of the LAHs can shed light on their origin, as well as, provide important information about the gas properties in the CGM of the fainter and smaller galaxies, which represent the bulk of the galaxy population at high redshift.

The paper is organized as follows: we describe our data and sample construction in Sect.~\ref{sec:2}. Section~\ref{sec:3} explains our binning and fitting procedures for the construction of spatially resolved maps of the $\lya$ distribution. Section~\ref{sec:4} describes the model that we use to determine the spectral characteristics of the LAHs presented in Sect.~\ref{sec:5}. Section~\ref{sec:5} also includes a statistical comparison with the central $\lya$ line properties. In Sect.~\ref{sec:6} we connect the spectral and spatial $\lya$ properties and investigate the link with the UV properties of the host galaxies. Finally, we discuss and present our summary and conclusions in Sect.~\ref{sec:7} and~\ref{sec:8}, respectively. 

In this paper, we use AB magnitudes, physical distances and assume a $\Lambda$CDM  cosmology  with $\Omega_{m}$ = 0.3, $\Omega_\Lambda$ = 0.7 and $H_{0}$ = 70 km s$^{-1}$ Mpc$^{-1}$.


\section{Data and sample definition}
\label{sec:2}

\subsection{Observations and data reduction}
\label{sec:21}

The MUSE UDF data were taken as part of the MUSE Guarantee Time Observations program between September 2014 and February 2016 under a clear sky, photometric conditions and good seeing (full width at half maximum, \mbox{FWHM}, of $0\farcs6$ at 7750\AA). More details about the data acquisition can be found in \cite{B17}.

Several $1\arcmin \times 1\arcmin$ pointings (corresponding to the MUSE field of view) were completed at two levels of depth: a $3\arcmin \times 3\arcmin$ medium-deep field consisting of a \mosaic\ of nine ten-hour pointings denoted \textsf{udf-0[1-9]} and an ultra-deep field, denoted \udft\, of a single $\approx$20 hour pointing that overlaps with the \mosaic\ reaching a total of $\approx$30 hours depth.

The data reduction of both the \udft\ and \mosaic\ data cubes has been improved with respect to the one we used in L17 (version 0.41) resulting in less systematics and an improved sky subtraction. Those latest data cubes (version 1.0) will be described in a forthcoming paper \cite{B19}. 

The data cubes contain $323\times322$ and $945\times947$ spatial pixels (spaxel) for the \udft\ and \mosaic\ field, respectively.
The spatial sampling is $0\farcs2\times0\farcs2$ and the spectral sampling is 1.25~\AA. 
The number of spectra matches the number of spatial pixels in each data cube with a wavelength range of 4750 \AA\ to 9350 \AA\ (3681 spectral pixels) and a wavelength-dependent spectral resolution ranging from $\approx$180 to $\approx$90 km\,s$^{-1}$ from blue to red (see Fig.~15 of \citealt{B17}).
The data cubes also contain the estimated variance for each pixel.
The data reaches a limiting surface brightness (SB) sensitivity (1$\sigma$) of 2.8 and 5.5$\times$10$^{-20}$ erg\,s$^{-1}$\,cm$^{-2}$ \AA$^{-1}$\,arcsec$^{-2}$ for an aperture of $1''\times1''$ in the 7000-8500 \AA\  range for the \udft\ and \mosaic\ data cubes, respectively (see \citealt{B19} for more details). 

Based on the reduced data cubes presented in \cite{B17}, \citeauthor{I17} (2017; hereafter I17) constructed the MUSE UDF catalog.
The source detection and extraction were performed using (i) priors detected in the Hubble Space Telescope (HST) data \citep{R15}, imposing a magnitude cut at 27 in the F775W band for the \mosaic\ field only, and (ii) the {\tt ORIGIN} detection software (\citeauthor{Ma17}, submitted). 
A complete description of the strategy used for the catalog construction can be found in I17.
{\tt The ORIGIN} software (see \citealt{B17} for technical details) is designed to detect emission lines in 3D data sets.
This software enabled the discovery of a large number of LAEs that were barely seen or even undetectable in the HST images \citep{M18}.

\subsection{Lyman alpha emitter sample}
\label{sec:22}

Our parent sample was constructed from the MUSE UDF LAH catalog (L17, Sect.~2.2). In this study, extended $\lya$ emission has been reported around 145 star-forming galaxies between the redshift range of 3 to 6.
From this sample of 145 LAEs, we removed 16 galaxies (i.e., 11\%) which have a double-peaked $\lya$ line profile. The study of such objects, which do not represent the bulk of the LAE population, is beyond the scope of this paper and will be subject of a future study. 
Next, we impose a signal-to-noise ratio (S/N) cut requiring a minimum value of 6 in the core component and in the halo component. The S/N of each component was calculated in L17 and can be found in Table~\ref{tab:0}. Below this S/N value the 1$\sigma$ errors on the resulting parameters of our fitting procedure (see Sect.~\ref{sec:41}) are very large, on the order of 100 km s$^{-1}$ and 250 km s$^{-1}$ on average for the peak position and width of the line, respectively.
Our final sample consists of 19 LAEs surrounded by a $\lya$ halo: 4 are in the \udft\ field and 15 are found only in the \mosaic\ field. 

\begin{figure*}
\centering
   \resizebox{\hsize}{!}{\includegraphics{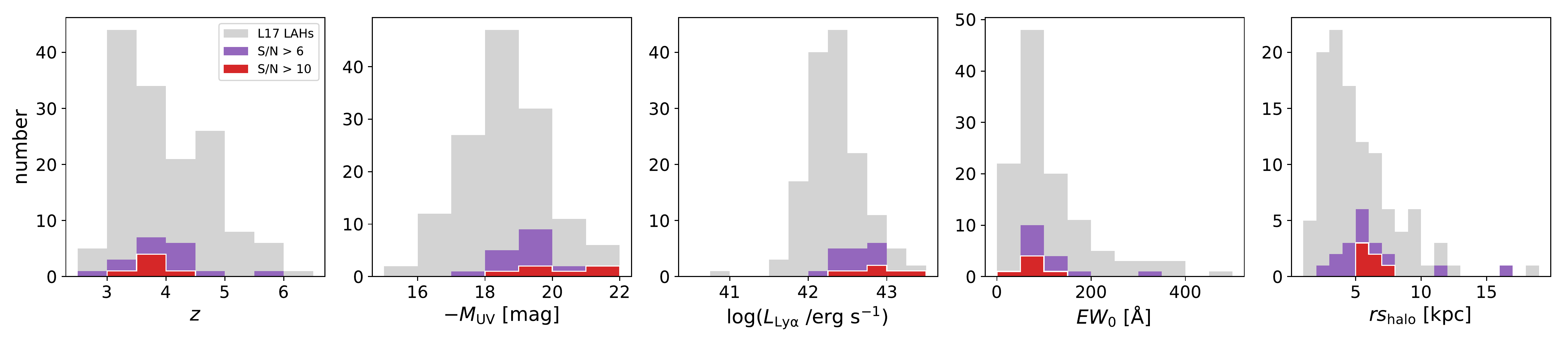}}
    \caption{\textit{From left to right:} Redshift, absolute UV magnitude, $\lya$ luminosity, $\lya$ rest-frame equivalent width, and $\lya$ halo scale length distributions of the parent sample (grey), the S/N > 6 (purple) and the S/N > 10 (red) samples.}
    \label{fig:0}
\end{figure*}

In this analysis, we use two samples with different S/N cuts: S/N > 6 and S/N > 10. The high S/N subsample consists of six LAEs and is used for our spatially resolved analysis (see Sect.~\ref{sec:3}).
The redshift, absolute UV magnitude, $\lya$ luminosity, $\lya$ rest-frame equivalent width, and $\lya$ halo scale length distributions of the parent sample (grey), the S/N > 6 (purple) and the S/N > 10 (red) samples are shown in Fig.~\ref{fig:0}. Those general properties are taken from the Table B.1 of L17 and are summarized in Table~\ref{tab:0}. The UV continuum slopes of our LAEs calculated in \citeauthor{H17}~(2017; hereafter H17) and the S/N values of the core and halo component calculated in L17 used for our sample selection are also indicated in this table.
In order to test the null hypothesis that our subsamples of LAHs are similar to the parent sample, we perform a Kolmogorov-Smirnov (KS) test. Our S/N > 6 sample has similar properties to that of the parent sample in terms of redshift (2.99 < $z$ < 5.98) and equivalent width (31.2 < EW$_0$[\AA] < 322.5) with KS statistic D$_{\rm KS}$ < 0.2 and corresponding p-value p$_0$ > 0.5. They are also statistically similar regarding their halo scale length distributions (2.1 < $rs_{\rm halo}$[kpc] < 16.4, D$_{\rm KS}$ = 0.33 and p$_0$ = 0.05). By construction, our S/N > 6 sample is brighter in $\lya$ emission ($1.5\times10^{42}$ < $L_{\rm \lya}$[erg s$^{-1}$] < $2.4\times10^{43}$) and UV continuum ($-$18 $\geq$ $M_{\rm UV}$ $\geq$ $-$21) with D$_{\rm KS}$ > 0.4 and p$_0$ < 0.006.

\begin{table*}
\caption{Global properties of our LAEs and their $\lya$ halo (taken from table B.1 of L17).}
\centering
\def\arraystretch{1.5}
\begin{tabular}{cccccccccc}
\hline
\hline
$\rm ID$ & $z$ & $-M_{\rm UV}$ & log$_{\rm 10}\ L_{\rm Ly\alpha}$ & $EW_{\rm 0}$ & $\beta$ & $rs_{\rm halo}$ & $rs_{\rm core}$ & S/N$_{\rm halo}$ & S/N$_{\rm core}$ \\
& & [AB] & [erg s$^{-1}$] & [\AA] & & [kpc] & [kpc] & & \\
\hline
82 & 3.6079 & 19.82 & 42.75 & 85.0 & -1.7$\pm$0.1 & 7.00$\pm$0.58 & 0.62$\pm$0.01 & 14.4 & 14.1 \\
149 & 3.7218 & 18.95 & 42.49 & 87.7 & -2.3$\pm$0.1 & 6.16$\pm$0.74 & 0.51$\pm$0.02 & 11.3 & 45.2 \\
180 & 3.4607 & 18.55 & 42.19 & 77.0 & -1.7$\pm$0.1 & 3.69$\pm$0.96 & 0.65$\pm$0.02 & 7.1 & 7.7 \\
547 & 5.9776 & 19.02 & 42.77 & 150.7 & -2.2$\pm$0.4 & 6.61$\pm$1.23 & 0.32$\pm$0.02 & 6.8 & 24.9 \\
1059 & 3.8063 & 21.32 & 43.09 & 52.8 & -1.2$\pm$0.0 & 6.87$\pm$0.45 & 0.47$\pm$0.00 & 18.1 & 42.0 \\
1113 & 3.0905 & 20.67 & 42.68 & 31.2 & -1.9$\pm$0.0 & 5.22$\pm$0.47 & 0.99$\pm$0.01 & 19.4 & 17.4 \\
1185 & 4.4996 & 21.17 & 43.38 & 97.9 & -1.9$\pm$0.2 & 5.73$\pm$0.21 & 1.35$\pm$0.00 & 39.2 & 35.5 \\
1283 & 4.3648 & 20.76 & 42.86 & 54.2 & -1.5$\pm$0.1 & 5.71$\pm$0.51 & 0.52$\pm$0.00 & 15.9 & 7.3 \\
1711 & 3.7662 & 19.25 & 42.47 & 71.8 & -1.8$\pm$0.2 & 5.90$\pm$1.42 & 0.59$\pm$0.02 & 6.5 & 14.3 \\
1723 & 3.6016 & 19.15 & 42.67 & 134.4 & -1.6$\pm$0.2 & 5.23$\pm$0.93 & 0.37$\pm$0.01 & 6.0 & 30.8 \\
1726 & 3.7075 & 19.2 & 42.79 & 145.9 & -2.0$\pm$0.2 & 5.41$\pm$0.59 & 0.25$\pm$0.01 & 12.6 & 26.2 \\
1761 & 4.0284 & 19.34 & 42.39 & 60.0 & -1.8$\pm$0.2 & 7.71$\pm$1.61 & 0.49$\pm$0.00 & 6.8 & 6.2 \\
1817 & 3.4155 & 18.86 & 42.55 & 120.8 & -2.0$\pm$0.1 & 4.45$\pm$1.30 & 0.45$\pm$0.01 & 6.1 & 17.5 \\
1950 & 4.4718 & 19.55 & 42.49 & 59.3 & -1.7$\pm$0.1 & 11.32$\pm$3.06 & 0.55$\pm$0.01 & 7.3 & 9.1 \\
2365 & 3.5975 & 18.18 & 42.91 & -- & -- & 4.77$\pm$0.81 & 0.28$\pm$0.01 & 6.7 & 10.9 \\
6416 & 4.2311 & 19.95 & 42.88 & -- & -- & 2.14$\pm$0.31 & 0.22$\pm$0.00 & 10.6 & 8.4 \\
6680 & 4.5046 & 18.86 & 42.7 & 139.8 & -2.3$\pm$0.2 & 16.43$\pm$8.78 & 0.87$\pm$0.02 & 7.0 & 105.6 \\
7047 & 4.2291 & 19.23 & 42.62 & 96.2 & -2.1$\pm$0.3 & 4.67$\pm$0.99 & 0.51$\pm$0.01 & 7.6 & 15.3 \\
7159 & 2.9958 & 17.63 & 42.49 & 322.5 & -2.1$\pm$0.4 & 3.80$\pm$0.80 & 0.29$\pm$0.02 & 7.3 & 15.1 \\
\end{tabular}
\tablefoot{ID: source identifier in the MUSE UDF catalog by I17. $z$: redshift in I17. $M_{\rm UV}$: absolute far-UV magnitude. log$_{\rm 10}\ L_{\rm Ly\alpha}$: logarithm of the $\lya$ luminosity in erg s$^{-1}$. $EW_{\rm 0}$: total $\lya$ rest-frame equivalent width in \AA. $\beta$: UV continuum slope calculated in H17. $rs_{\rm halo}$: exponential scale length of the $\lya $ halo in physical kpc. $rs_{\rm core}$: exponential scale length of the UV continuum in physical kpc. S/N$_{\rm halo}$: signal-to-noise ratio of the $\lya $ halo. S/N$_{\rm core}$: signal-to-noise ratio of the $\lya $ core (i.e., “continuum-like”) component.}

\label{tab:0}
\end{table*}


\section{Resolved spectroscopy of the $\lya$ haloes}
\label{sec:3}

In order to construct detailed and reliable maps of the diffuse $\lya$ emission properties around galaxies, we need objects with relatively bright $\lya$ haloes.
We therefore run our binning and fitting procedures on the six galaxies of the high S/N subsample.
In this spectral study of the $\lya$ haloes, the measured parameters are the peak velocity shift ($\Delta v$) relative to the central ($r$ < 0\farcs4) line, the \mbox{FWHM} and the asymmetry parameter ($a_{\rm asym}$) of the $\lya$ line.

\subsection{Data binning}
\label{sec:31}

In order to reveal the spatial variations of the $\lya$ line profiles within the extended gaseous haloes, we construct two-dimensional (2D) binned resolved maps. This method allows us to increase the S/N in the outermost parts of the LAH where the surface brightness drops significantly.

The 2D binning is performed on the $\lya$ NB image constructed from the MUSE data cube using the Voronoi tessellation method introduced in \cite{C03}.   
The spectral window of the NB images varies from source to source and is defined to include the total $\lya$ line within a spatial aperture of radius $r_{\rm CoG}$ determined by using a curve of growth (CoG) method\footnote{The CoG method is applied on the $\lya$ NB image optimized in S/N as defined in L17.}. The borders of the line are set when the flux goes below zero. The spectral bandwidth is then widened by few angstroms on each side to ease the fit procedure (see Sect.~\ref{sec:32}). This method allows us to encompass all the $\lya$ flux while limiting the noise.

We choose a target S/N of 6 for the binning meaning that the spaxels above this threshold are not combined with others. We note that the S/N values of the bins resulting from the 2D Voronoi binning method (see \citealt{C03}) are symmetrically clustered around the target S/N, value of 6 here. This explains why some bins have a S/N lower than 6.

\subsection{Line extraction and measurements}
\label{sec:32}

We extract the $\lya$ line in each resulting Voronoi bin and measure its properties by fitting an asymmetric Gaussian function. The line model used was introduced by \cite{S14} and is expressed by: 
\begin{equation}
\label{eq:gaussasym}
f(\lambda) = A \exp\left( \dfrac{- (\lambda-\lambda_0)^2}{2\ (a_{\rm asym}(\lambda-\lambda_0)+d)^2}\right), 
\end{equation}
where A, $\lambda_0$, $a_{\rm asym}$ and $d$ are the amplitude, the peak wavelength, the asymmetry parameter and a typical width of the line, respectively. \cite{S14} argue that Eq.~(\ref{eq:gaussasym}) provides a more  robust peak position for the typical LAE profiles than a symmetric Gaussian profile. From Eq.~(\ref{eq:gaussasym}), we derive the analytic expression for the full width at half maximum (not corrected for instrumental effects) of the line as:
\begin{equation}
\mbox{FWHM} = \dfrac{2\sqrt{2\ln2} \ d}{1-2\ln2 \ a_{\rm asym} ^2}.
\end{equation}

In each bin, we fit the $\lya$ line and its associated variance using the Python package EMCEE \citep{Fo13} as our Markov chain Monte Carlo (MCMC) sampler to determine the joint likelihood of our parameters in Eq.~(\ref{eq:gaussasym}). We use the measured parameters of the $\lya$ line integrated in an aperture of radius $r_{\rm CoG}$ as priors for the parameter space exploration.

To perform the fit, we use 50 walkers and run the MCMC for 5000 steps for each Voronoi bin discarding the initial 1000 steps. 
We use the median values of the resulting posterior probability distributions for all the model parameters. The  errors  on  the parameters  are  estimated  using  the  16$^{\rm th}$ and 84$^{\rm th}$ percentiles.

We check the lines and the corresponding fits in each Voronoi bin. A visual inspection per bin shows that below a S/N of 4, the quality of the fit in the bin is poor. It is confirmed when looking at the uncertainties of the fit parameters. As a consequence, we reject the bins whose S/N is lower than 4. 
The selected bins are located in areas where the surface brightness is brighter than 10$^{-18}$ $\sbl$. We also check the reduced $\chi^{2}$ and found that it is < 1.5 and > 0.1 for all the S/N > 4 fitted lines and < 1 for more than 90\% of the spectra for each object. This result confirms that the $\lya$ lines are fairly well described by an asymmetric Gaussian function.

\subsection{Resolved $\lya$ spectral properties}
\label{sec:33}

\begin{figure*}
\centering
   \resizebox{.98\hsize}{!}{\includegraphics{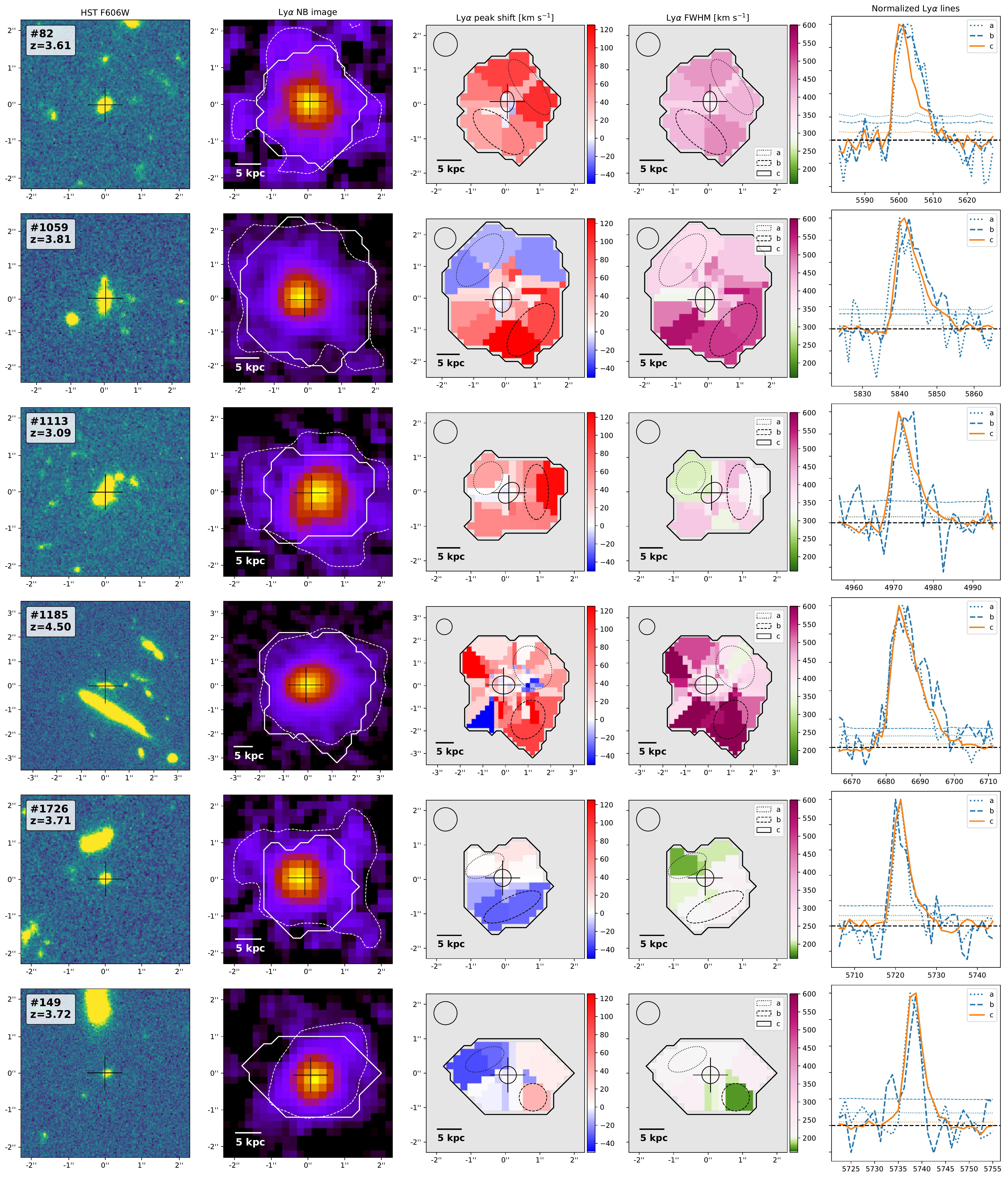}}
    \caption{Sample of six LAEs surrounded by a $\lya$ halo with S/N > 10 from the L17 sample. Each row shows a different object. 
    \textit{First column:} HST/F606W image of the LAE. The HST coordinates \citep{R15} are indicated by the black cross in all panels. The MUSE ID and $z$ are indicated.
    \textit{Second column:} $\lya$ narrowband image (plotted with a power-law stretch) with SB contour at 10$^{-18} \ \sbl$ (dashed white). The solid white contour shows the outer limit of the Voronoi bins group (S/N > 4) used in this study (see Sect.~\ref{sec:31}). 
    \textit{Third column:} Map of the $\lya$ line peak velocity shift relative to the central (r < $0\farcs4$) $\lya$ line peak (see Sect.~\ref{sec:33}). 
    The diverging colormaps are centered on the parameter values of the central line and have the same dynamical range to ease the visual comparison.
    The \mbox{FWHM} of the PSF is plotted in the upper-left corner.
    \textit{Fourth column:} Map of the $\lya$ line \mbox{FWHM}. 
    \textit{Fifth column:} Lines extracted in the ellipsoid areas designated by the same line style on the maps. They aim to highlight the spectral variations in the core and halo components. The dotted and dashed colored horizontal lines show the 1$\sigma$ error and the dashed black line shows the zero.}
    \label{fig:1}
\end{figure*}

The $\lya$ emission maps of the six galaxies which meet our S/N criterion are shown in Fig.~\ref{fig:1}. 
By requiring a S/N > 4 in the bins (see Sect.~\ref{sec:32}), our maps consist of 14, 41, 23, 106, 16 and 21 bins for object \#82, \#1059, \#1113, \#1185, \#1726 and \#149, respectively, probing the LAH in a radius of $\approx10-15$ kpc. 
In order to facilitate the visual comparison between the central and the halo components of the galaxies, the diverging colormaps of the peak velocity shift and \mbox{FWHM} maps (third and fourth columns, respectively) are centered on the best-fit parameter values of the $\lya$ line extracted in the central region (aperture of radius 0.4$''$ or 2 MUSE pixels) of the galaxy. By construction, the central best-fit peak value is very close (< 25 km s$^{-1}$ difference) to the position corresponding to the $\lya$ redshift from the I17 catalog where a point spread function (PSF) weighted extraction has been used. The last column in Fig.~\ref{fig:1} shows the $\lya$ lines extracted in three different and arbitrarily chosen regions, one is located in the inner region of the galaxy (solid orange ellipse) and the others in the surrounding halo (dotted and dashed blue ellipses).

It is apparent from Fig.~\ref{fig:1} that the LAEs exhibit significant differences in their line profiles as a function of spatial position.

\begin{figure*}
    \centering
    \begin{subfigure}[t]{\hsize}
        \centering
        \includegraphics[width=\hsize]{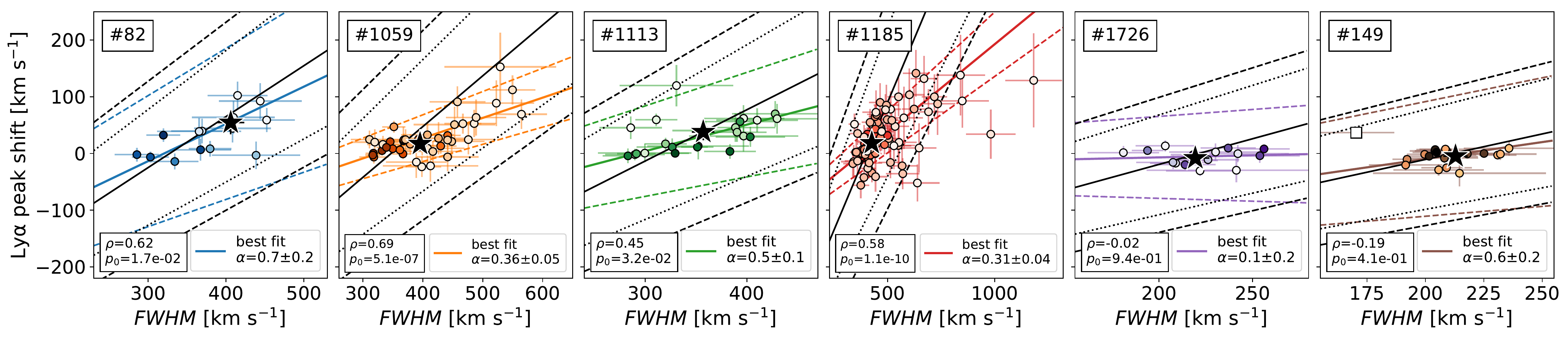}
        \label{fig:2a}
    \end{subfigure}%
    \\
	\bigskip
    \begin{subfigure}[t]{\hsize}
        \centering
        \includegraphics[width=\hsize]{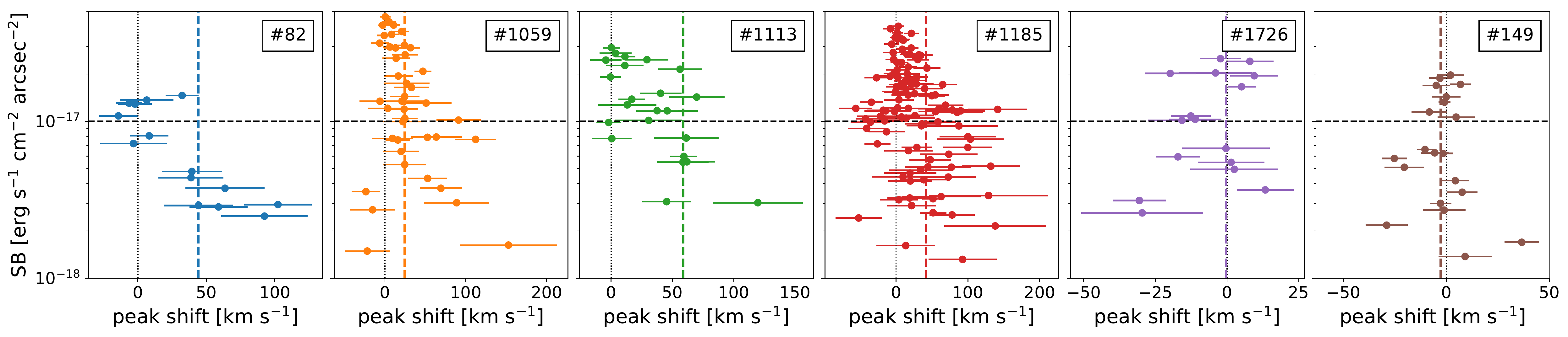}
    \end{subfigure}
    \begin{subfigure}[t]{\hsize}
        \centering
        \includegraphics[width=\hsize]{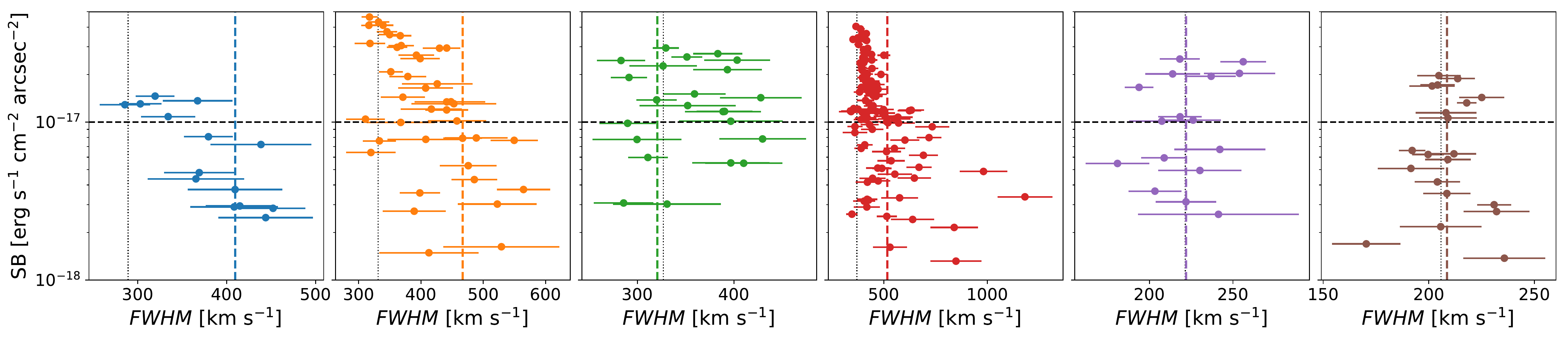}
        \label{fig:2b}
    \end{subfigure}
    \caption{$\lya$ haloes spectral analysis of the six galaxies presented in Fig.~\ref{fig:1}. 
    	\textit{Top row} : Peak velocity shift relative to the central $\lya$ line (r < $0\farcs4$) plotted against the \mbox{FWHM} of the $\lya$ lines extracted in the different Voronoi bins for the six tested objects (MUSE IDs are indicated). 
        The brightness of the points indicates the bin surface brightness (darker point meaning higher SB).
        Pearson correlation coefficients $\rho$ and corresponding $p_{\rm 0}$ values are shown in each panel. The solid colored line indicates the best fit of our data, while the dashed colored lines show the 1$\sigma$ errors (the slope $\alpha$ and its error are indicated in the legend). The square point (last panel) is discarded for the fit (sigma clipping factor of 4).
        The parameters measured on the total $\lya$ lines are indicated by the star symbols. 
        For comparison the re-scaled V18 relation (solid line) and its dispersion (dotted lines) are plotted in black. The black dashed lines show the V18 relation 1$\sigma$ errors by taking into account the errors of the re-scaling procedure (see Sect.~\ref{sec:332}).
    	\textit{Bottom rows} : $\lya$ surface brightness of the bins as a function of the peak velocity shift (top) and \mbox{FWHM} (bottom) of the $\lya$ line for the six tested objects. The black dotted vertical lines indicate the central $\lya$ line values. The black horizontal dashed line represents the arbitrary SB threshold at 10$^{-17}$ erg s$^{-1}$ cm$^{-2}$ arcsec$^{-2}$. The colored vertical dashed lines show the median values of the SB < 10$^{-17}$ erg s$^{-1}$ cm$^{-2}$ arcsec$^{-2}$ bins.
    	}
	\label{fig:2}
\end{figure*}

\subsubsection{$\lya$ line properties}
\label{sec:331}

The resolved maps of the $\lya$ peak velocity shift (third column) reveal that the peak position does vary. The amplitudes of those variations are lower than $\approx$200 km s$^{-1}$ for the six tested objects. 
We do not observe any obvious velocity field in the $\lya$ distributions except for the object \#1059 (second row) and possibly object \#1726 (second to last row) where a structured velocity field is suggested (see discussion in Sect.~\ref{sec:72}).

The resolved maps of the $\lya$ \mbox{FWHM} (fourth column) reveal that the width of the $\lya$ line varies spatially as well. The amplitude of those variations can be higher, notably for the objects \#1059 and \#1185 for which the line width variations reach $\approx$ 250 and > 500 km s$^{-1}$, respectively.
 
Generally, a large diversity in the line profiles is observed: while some galaxies globally show a wider and redder $\lya$ line in the outer regions (e.g., objects \#82, \#1059 and  \#1113), others show slightly narrower and bluer lines (e.g., objects \#1726 and \#149). 
One can also observe spectral variations at small spatial scales (< 5 kpc) within the LAHs.
The last column of Fig.~\ref{fig:1} illustrates some of those spectral changes by showing three spectra integrated in three arbitrarily chosen regions.
Interestingly, the $\lya$ line extracted in the bins are all single-peaked except for object \#149 where a blue bump is visible (dashed line) although its S/N is low (see discussion Sect.~\ref{sec:733}).

\subsubsection{Peak velocity shift versus width of the $\lya$ line}
\label{sec:332}
The top row of Fig.~\ref{fig:2} displays the relation between the best-fit parameters of the $\lya$ lines extracted in the different bins for the six tested objects in terms of peak velocity shift and \mbox{FWHM}. The points are color coded by the SB of the bins (darker point meaning higher SB). 

We find a significant correlation between the $\lya$ peak velocity shift and the \mbox{FWHM} for four objects out of six (see the Pearson coefficients). This result is in good agreement with C19 which also found this correlation for spatially resolved spectra of two lensed galaxies. For \#1726 and \#149, the small dynamic range in \mbox{FWHM} and peak velocity shift precludes any statement about the lack of existence of a correlation with the present data.

We compare our results to the empirical relation established in \citeauthor{V18}~(2018; hereafter V18) between the $\lya$ peak velocity shift relative to the systemic redshift and the \mbox{FWHM} of the $\lya$ line. This object-by-object based relation has been established using 13 LAEs detected in various MUSE fields for which the systemic redshift is known as well as spectroscopic $\lya$ data found in the literature spanning a wide redshift range. 
We performed a vertical re-scaling of the V18 relation (using a least-squares minimization method) in order to aid the visual comparison of the slopes. This is due to the fact that we do not know the systemic redshift of our objects.
The re-scaled V18 relation (solid black line) and its dispersion (dotted black lines) are superimposed on our data in Fig.~\ref{fig:2} (top row). 
Remarkably, most of our data points fall close to this relation and within its 1$\sigma$ dispersion. The dashed black lines represent the V18 dispersion when including the errors from our re-scaling procedure.
The colored solid and dashed lines show our best fit to the data and the 1$\sigma$ errors, respectively. Similarly to V18, we run the {\tt LTS\_LINEFIT} routine\footnote{\cite{LTS14} and \url{http://www-astro.physics.ox.ac.uk/\~mxc/software/\#lts}} introduced in \cite{C13} which uses a robust least-squares fitting technique and takes into account the errors of both variables. A sigma-clipping factor of 4 is used in the fit procedure. The excluded point is shown by the square symbol (last panel).

We find a trend similar as V18 for spatially resolved spectra but the slope of the relation is lower and varies from one object to another (see values on Fig.~\ref{fig:2}, top row). 
Interestingly, the best-fit parameters measured on the total $\lya$ lines (extracted in an aperture of radius $r_{\rm CoG}$) of the six objects (black star symbols) are well located on the V18 relation.

\subsubsection{Surface brightness effects}
\label{sec:333}

The bottom rows of Fig.~\ref{fig:2} shows the surface brightness in the bins as a function of the $\lya$ peak velocity shift relative to the $\lya$ central line (top panels) and the \mbox{FWHM} of the line in the Voronoi bins (bottom panels).
In a similar way to C19, the vertical dashed lines show the median values of the faintest bins (SB < 10$^{-17}$ erg s$^{-1}$ cm$^{-2}$ arcsec$^{-2}$, horizontal dashed line).

On average, the $\lya$ lines extracted from the brightest bins (SB > 10$^{-17}$ erg s$^{-1}$ cm$^{-2}$ arcsec$^{-2}$) have smaller velocity offset (< 50 km s$^{-1}$) and are narrower compared to the ones extracted in the faintest bins for which we observe more scatter.
This result is in good agreement with C19 (see their Fig.~4). 
We remark that the bins of low SB are also, by construction, the more spatially extended and thus more prone to artificial broadening (see Sect.~\ref{sec:72} for discussion).

This statement does not apply to object \#1726 and \#149 for which the variations are smaller. The object \#1113 is also interesting as we find broad $\lya$ lines in both faint and bright bins.\\
\newline 
All in all, even at the MUSE spectral and spatial resolution, we detect significant variations of the $\lya$ line profile within the spatial $\lya$ distribution, for this sample of six bright LAHs with high S/N.


\section{Ly$\alpha$ halo 3D decomposition}
\label{sec:4}

In order to test the robustness of the trends found in the first part of the paper on a larger sample, we set up a new statistical method based on the W16 and L17 analyses which consists of the simultaneous 3D (i.e., spectral and spatial) decomposition of the $\lya$ emission. This parametric fit procedure is less sensitive to the noise. It therefore allows us to increase the sample by including lower S/N $\lya$ haloes (S/N > 6).

\begin{table*}
\caption{Fitting results from our 3D two-component analysis.}
\centering
\def\arraystretch{1.5}
\begin{tabular}{cccccccc}
\hline
\hline
$\rm ID$ & FWHM$_{\rm CORE}$ & FWHM$_{\rm HALO}$ & $a_{\rm asym,CORE}$ & $a_{\rm asym,HALO}$ & $\lambda \rm peak_{\rm CORE}$ & $\lambda \rm peak_{\rm HALO}$ & $\Delta \rm v_{\rm HALO-CORE}$ \\
& [km s$^{-1}$] & [km s$^{-1}$] &  &  & [\AA\ ] & [\AA\ ] & [km s$^{-1}$] \\
\hline
82 & 121$_{-28}^{+37}$ & 419$_{-29}^{+30}$ & 0.32$_{-0.03}^{+0.02}$ & 0.05$_{-0.05}^{+0.05}$ & 5599.9$_{-0.2}^{+0.2}$ & 5603.7$_{-0.4}^{+0.4}$ & 204$_{-32}^{+32}$ \\
149 & 55$_{-8}^{+10}$ & 37$_{-16}^{+27}$ & 0.29$_{-0.01}^{+0.01}$ & 0.04$_{-0.17}^{+0.17}$ & 5737.9$_{-0.0}^{+0.0}$ & 5737.8$_{-0.3}^{+0.3}$ & -4$_{-20}^{+17}$ \\
180 & 326$_{-41}^{+37}$ & 294$_{-110}^{+89}$ & 0.14$_{-0.08}^{+0.08}$ & 0.23$_{-0.17}^{+0.13}$ & 5421.2$_{-0.7}^{+0.6}$ & 5420.3$_{-0.9}^{+1.5}$ & -47$_{-88}^{+115}$ \\
547 & 147$_{-21}^{+22}$ & 168$_{-82}^{+105}$ & 0.37$_{-0.03}^{+0.03}$ & 0.19$_{-0.27}^{+0.18}$ & 8479.5$_{-0.2}^{+0.2}$ & 8480.0$_{-0.8}^{+1.0}$ & 17$_{-36}^{+42}$ \\
1059 & 224$_{-18}^{+19}$ & 420$_{-29}^{+28}$ & 0.32$_{-0.02}^{+0.02}$ & 0.18$_{-0.06}^{+0.06}$ & 5840.5$_{-0.1}^{+0.1}$ & 5842.1$_{-0.5}^{+0.5}$ & 80$_{-31}^{+31}$ \\
1113 & 82$_{-26}^{+42}$ & 326$_{-31}^{+30}$ & 0.32$_{-0.01}^{+0.01}$ & 0.12$_{-0.06}^{+0.06}$ & 4970.6$_{-0.1}^{+0.2}$ & 4973.1$_{-0.4}^{+0.4}$ & 150$_{-32}^{+35}$ \\
1185 & 278$_{-14}^{+14}$ & 471$_{-16}^{+15}$ & 0.30$_{-0.02}^{+0.01}$ & 0.12$_{-0.03}^{+0.03}$ & 6683.4$_{-0.1}^{+0.1}$ & 6685.6$_{-0.3}^{+0.3}$ & 99$_{-18}^{+18}$ \\
1283 & 192$_{-82}^{+95}$ & 458$_{-66}^{+65}$ & 0.39$_{-0.05}^{+0.05}$ & 0.36$_{-0.10}^{+0.09}$ & 6518.5$_{-0.3}^{+0.4}$ & 6521.1$_{-0.7}^{+0.8}$ & 118$_{-48}^{+55}$ \\
1711 & 220$_{-39}^{+31}$ & 146$_{-114}^{+99}$ & 0.23$_{-0.06}^{+0.06}$ & 0.18$_{-0.20}^{+0.12}$ & 5792.2$_{-0.3}^{+0.4}$ & 5791.5$_{-0.6}^{+0.7}$ & -40$_{-50}^{+56}$ \\
1723 & 156$_{-16}^{+18}$ & 295$_{-98}^{+112}$ & 0.26$_{-0.02}^{+0.02}$ & -0.20$_{-0.18}^{+0.34}$ & 5591.8$_{-0.1}^{+0.1}$ & 5594.2$_{-1.5}^{+1.1}$ & 129$_{-84}^{+62}$ \\
1726 & 103$_{-25}^{+25}$ & 45$_{-16}^{+31}$ & 0.30$_{-0.02}^{+0.02}$ & 0.07$_{-0.18}^{+0.21}$ & 5720.5$_{-0.1}^{+0.1}$ & 5720.8$_{-0.4}^{+0.3}$ & 12$_{-25}^{+20}$ \\
1761 & 104$_{-64}^{+101}$ & 252$_{-101}^{+207}$ & 0.38$_{-0.12}^{+0.08}$ & 0.21$_{-0.47}^{+0.33}$ & 6109.9$_{-0.6}^{+0.6}$ & 6111.3$_{-1.2}^{+1.7}$ & 67$_{-89}^{+114}$ \\
1817 & 135$_{-21}^{+22}$ & 33$_{-17}^{+34}$ & 0.27$_{-0.03}^{+0.02}$ & 0.08$_{-0.20}^{+0.13}$ & 5366.0$_{-0.1}^{+0.1}$ & 5365.9$_{-0.3}^{+0.4}$ & -2$_{-25}^{+30}$ \\
1950 & 240$_{-52}^{+48}$ & 285$_{-99}^{+122}$ & -0.15$_{-0.17}^{+0.18}$ & 0.14$_{-0.35}^{+0.21}$ & 6650.8$_{-0.9}^{+0.9}$ & 6648.9$_{-1.2}^{+1.9}$ & -84$_{-98}^{+123}$ \\
2365 & 153$_{-37}^{+38}$ & 178$_{-63}^{+80}$ & 0.32$_{-0.06}^{+0.05}$ & 0.25$_{-0.16}^{+0.10}$ & 5586.9$_{-0.2}^{+0.2}$ & 5588.0$_{-0.5}^{+0.6}$ & 64$_{-39}^{+46}$ \\
6416 & 294$_{-25}^{+24}$ & 115$_{-20}^{+24}$ & 0.12$_{-0.04}^{+0.04}$ & 0.27$_{-0.04}^{+0.03}$ & 6359.1$_{-0.4}^{+0.4}$ & 6356.6$_{-0.2}^{+0.2}$ & -115$_{-27}^{+26}$ \\
6680 & 106$_{-6}^{+6}$ & 70$_{-41}^{+76}$ & 0.24$_{-0.01}^{+0.01}$ & 0.29$_{-0.20}^{+0.07}$ & 6689.3$_{-0.0}^{+0.0}$ & 6688.8$_{-0.6}^{+0.6}$ & -22$_{-28}^{+27}$ \\
7047 & 143$_{-48}^{+42}$ & 141$_{-75}^{+71}$ & 0.31$_{-0.05}^{+0.04}$ & 0.20$_{-0.18}^{+0.11}$ & 6355.0$_{-0.2}^{+0.3}$ & 6355.7$_{-0.5}^{+0.9}$ & 37$_{-36}^{+53}$ \\
7159 & 40$_{-18}^{+18}$ & 331$_{-85}^{+83}$ & 0.01$_{-0.14}^{+0.16}$ & -0.15$_{-0.11}^{+0.11}$ & 4856.4$_{-0.2}^{+0.2}$ & 4856.2$_{-0.7}^{+0.6}$ & -12$_{-56}^{+51}$ \\
\end{tabular}
\tablefoot{ID: source identifier in the MUSE UDF catalog by I17. FWHM$_{\rm CORE}$: rest-frame full width at half maximum of the $\lya$ line extracted in the core component in km s$^{-1}$. FWHM$_{\rm HALO}$: rest-frame full width at half maximum of the $\lya$ line extracted in the halo component in km s$^{-1}$. $a_{\rm asym,CORE}$: asymmetry parameter of the $\lya$ line in the core. $a_{\rm asym,HALO}$: asymmetry parameter of the $\lya$ line in the halo. $\lambda \rm peak_{\rm CORE}$: peak wavelength position of the $\lya$ line in the core in \AA. $\lambda \rm peak_{\rm HALO}$: peak wavelength position of the $\lya$ line in the halo in \AA. $\Delta \rm v_{\rm HALO-CORE}$: halo/core peak separation in km s$^{-1}$. Those values are corrected for the MUSE LSF (see Table~\ref{tab:1b} for non-corrected values).}

\label{tab:1}
\end{table*}

\begin{figure*}
\centering
   \resizebox{\hsize}{!}{\includegraphics{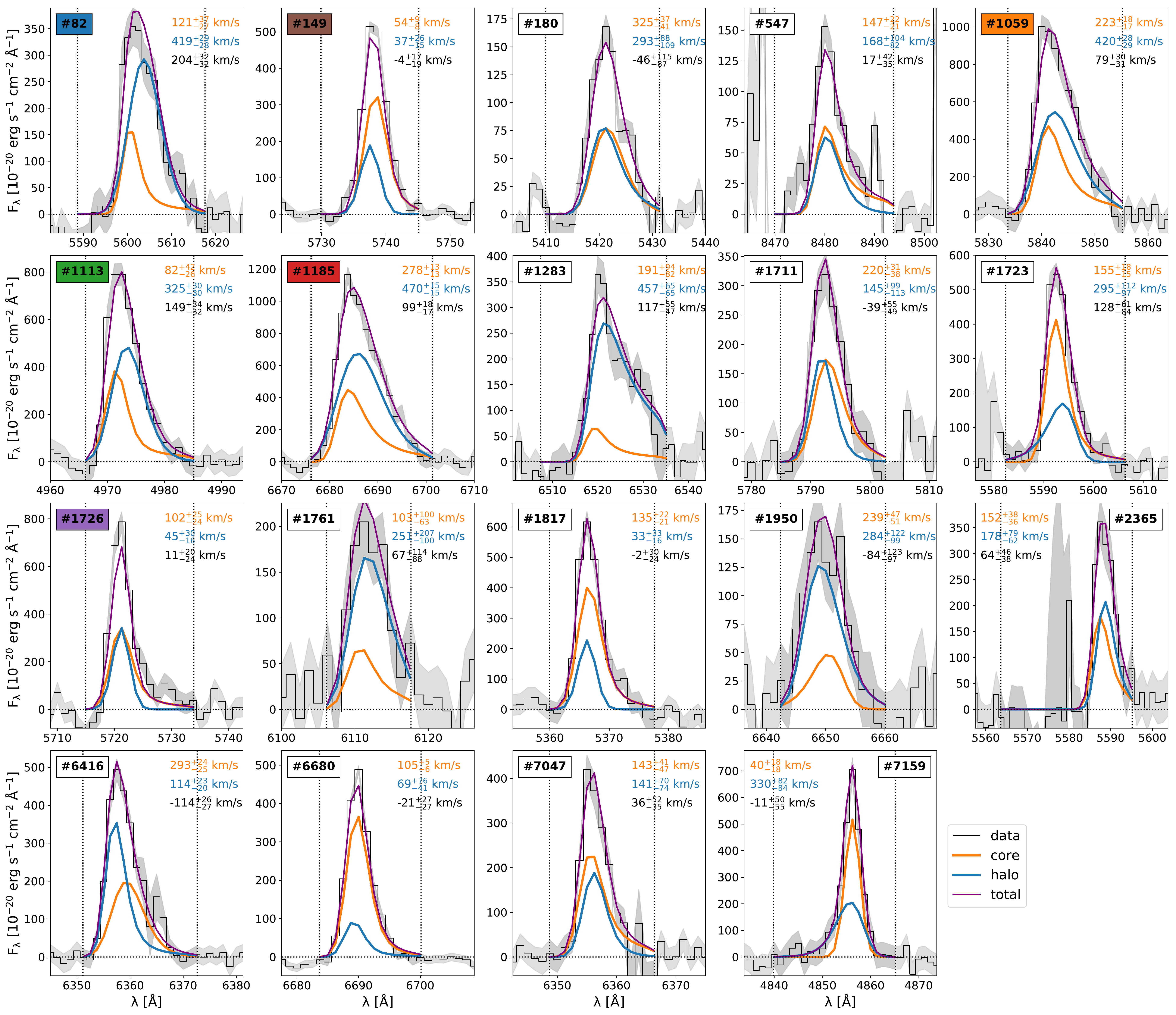}}
    \caption{1D profiles of the 3D modeled $\lya$ distribution decomposed into central (orange lines) and extended (blue lines) components (see Sect.~\ref{sec:41} and Eq.~(\ref{eq:model})). These represent the 19 galaxies for which the S/N in the core and halo components is higher than 6. The MUSE identifier are indicated in each panel. The colored labels correspond to the S/N > 10 objects where the same color coding as in Fig.~\ref{fig:2} is used. The total spectra of the modeled $\lya$ emission are shown in purple. For comparison, the observed $\lya$ lines and their 1$\sigma$ errors (see Sect.~\ref{sec:412}) are overplotted in black and grey respectively. The vertical dotted black lines delimit the spectral window used for the fit (see Sect.~\ref{sec:412}). The best-fit line parameters (halo/core \mbox{FWHM} and peak separation) are indicated in km\,s$^{-1}$ (blue, orange and black, respectively) in each panel.}
    \label{fig:3}
\end{figure*}

\subsection{Three-dimensional two-component fits}
\label{sec:41}

Our previous studies (W16 and L17) have demonstrated that the spatial distribution of $\lya$ radiation emitted from and around distant star-forming galaxies is well-described by a two-dimensional, two-component circular exponential distribution. 
In order to push the analysis further, we upgrade the model by adding the spectral dimension to the model. We thus use a 3D, two-component modeling approach, fitting MUSE $\lya$ subcubes. This method aims to disentangle and compare the averaged $\lya$ spectral signatures of the host galaxy and its surrounding gaseous envelop. 

\subsubsection{Model}
\label{sec:411}

We describe here how we characterize the 3D $\lya$ distribution of our extended LAEs. 
In W16 and L17 we decomposed the observed spatial $\lya$ distribution into a central and an extended circular and co-spatial exponential component using the HST morphological information as prior (see Sect.~4.1 of L17). Using this method, we parametrized the spatial distribution of the $\lya$ emission with four parameters: the scale lengths and central flux intensities of the central and extended components ($rs_{\rm core}$, $rs_{\rm halo}$, $I_{\rm core}$, and $I_{\rm halo}$, respectively). 
Our 3D model is based on this 2D, two-component, circularly symmetric model of L17 and additionally takes into account the spectral dimension. 
It is defined by Eq.~(\ref{eq:model}) where C($r,\lambda$) describes the total $\lya$ flux distribution in the subcube of the source:
\begin{multline} 
    \label{eq:model} 
		C(r,\lambda) = \left[ I_{\rm core}(\lambda) \ \exp(-\frac{r}{rs_{\rm core}}) + 		I_{\rm halo}(\lambda) \ \exp\left( -\frac{r}{rs_{\rm halo}}\right) \right]  \\ 
		\times \ \rm PSF_{\rm 3D, MUSE}(\lambda_{\rm Ly\alpha}),
\end{multline}
The central flux intensity parameters $I_{\rm core}$ and $I_{\rm halo}$ depend on the wavelength and are described by the asymmetric Gaussian function given by Eq.~(\ref{eq:gaussasym}). 
Our fit procedure takes into account the MUSE 3D PSF so that the spatial and spectral parameters are corrected from the spatial PSF and line spread function (LSF) effects, respectively.

\subsubsection{Subcubes construction}
\label{sec:412}

From the MUSE data cubes, we extract subcubes centered on the spectral and spatial flux maxima of our objects. The spatial and spectral apertures used are different for each source. The radius of the spatial aperture corresponds to the measured CoG radius (see Sect. 5.3.2. of L17). The spectral window corresponds to the $\lya$ full line width extracted in the aperture of radius $r_{\rm CoG}$ expanded by 2.5 \AA\ (i.e., 2 MUSE pixels) on each side of the line. This 3D aperture ensures that most of the detectable $\lya$ flux is encompassed for each object while limiting noise. The continuum is removed by performing a spectral median filtering on the MUSE cubes (see Sect. 3.1.1 of L17 for more details). 

Following this procedure, we construct $\lya$ subcubes with spatial and spectral apertures ranging from 1\farcs8 to 4\farcs6  (median value of 2\farcs8) in radius and 11.25 \AA\  to 31.25 \AA\  (median value of 21.25 \AA\ ), respectively. This represents typically a total spectral windows of 1100 km s$^{-1}$.

\subsubsection{Fitting procedure}
\label{sec:413}

The modeling is performed by fitting the subcube of our sources by (i) fixing the core and halo scale length parameters $rs_{\rm core}$ and $rs_{\rm halo}$ to the values measured in L17 -- this step reduces the numbers of fitted parameters and therefore makes the fit more robust --, (ii) taking into account the 3D PSF by convolving the model with the MUSE spatial PSF similarly to L17 and the MUSE LSF, and (iii) making use of the variance of each 3D pixel of the subcube.

We thus have eight parameters in total to fit the $\lya$ 3D distribution: the amplitudes, the peak wavelengths, the asymmetry parameters and the full widths at half maximum of the two asymmetric Gaussian functions describing the flux intensity in the core and halo components as a function of wavelength.

To perform a robust fit of our MUSE subcubes and similarly to the 1D spectral fit performed in the first part of the paper (see Sect.~\ref{sec:32}), we use the Python package EMCEE \citep{Fo13} as our MCMC sampler to determine the joint likelihood of our parameters in Eq.~(\ref{eq:model}). We start by initializing the walkers with the following priors on the parameters: 
\begin{itemize}
\item The total flux of the lines in the core and halo are equal and are two times smaller than the total $\lya$ line integrated in an aperture of radius $r_{\rm CoG}$,
\item the wavelength of the core/halo line peaks are equal and correspond to the peak position of the total $\lya$ line,
\item \mbox{FWHM}$_{\rm CORE}$ and \mbox{FWHM}$_{\rm HALO}$ are equal and correspond to the \mbox{FWHM} of the total $\lya$ line,
\item the lines are symmetrical ($a_{\rm asym}$ = 0).
\end{itemize}
We use 100 walkers and run the MCMC for 5000 steps for each subcubes discarding the initial 1000 steps.
We use the median values of the resulting posterior probability distributions for all the model parameters. The  errors  on  the parameters  are  estimated  using  the  16$^{\rm th}$ and 84$^{\rm th}$ percentiles. The fitting results are given in Table~\ref{tab:1}.

Figure~\ref{fig:3} shows the spectral projection of the best-fit 3D model obtained for our 19 LAEs (S/N > 6 sample). We show the spectral decomposition of the total $\lya$ line (purple) into the core $\lya$ line (orange) and the halo $\lya$ line (blue).
We overplot the total observed $\lya$ line (black line with the 1$\sigma$ errors shown by the grey area) integrated in the MUSE subcubes (see previous section). For most of our objects, the modeled line is a good fit to the observed $\lya$ line. The adopted model thus appears to be a good representation of the observed data.

\subsection{Model robustness}
\label{sec:42}

In order to test our model, we compare the resulting 3D best-fit parameters to the binned resolved $\lya$ maps (see Fig.~\ref{fig:1}) obtained in the first part of the paper (Sect.~\ref{sec:3}). Because the line parameters of the resolved maps are not corrected for instrumental effects, we consider the 3D best-fit parameters not corrected for the LSF (presented in Table~\ref{tab:1b}) for this exercise.

As a first step, a qualitative visual inspection shows that the trends are similar for the six tested galaxies: 
\begin{itemize}
\item
In Fig.~\ref{fig:1} the objects \#82, \#1059 and \#1185 show broader and redder $\lya$ lines in the outermost regions of the $\lya$ spatial distribution. It is in good agreement with the 3D fit parameters. 
\item
On the contrary, the maps of the objects \#1726 and \#149 indicate that the $\lya$ lines of the outer bins are bluer and narrower compared to the ones of the inner bins, although the variations are small (< 30 km\,s$^{-1}$). It again agrees with the 3D fit results. 
\item
The resolved maps of the object \#1059 are more contrasted in the halo. The shift of the peak goes from $\approx-50$ km s$^{-1}$ to $\approx+150$ km s$^{-1}$ and the \mbox{FWHM} varies significantly throughout the halo. In this case the 3D fit provides information about the average $\lya$ line and as a consequence conceals the spectral variations (see the discussion sect.~\ref{sec:72}). Considering this effect, the line parameters resulting from the two methods are in rather good agreement.
\end{itemize}

\begin{figure}
\centering
   \resizebox{.7\hsize}{!}{\includegraphics{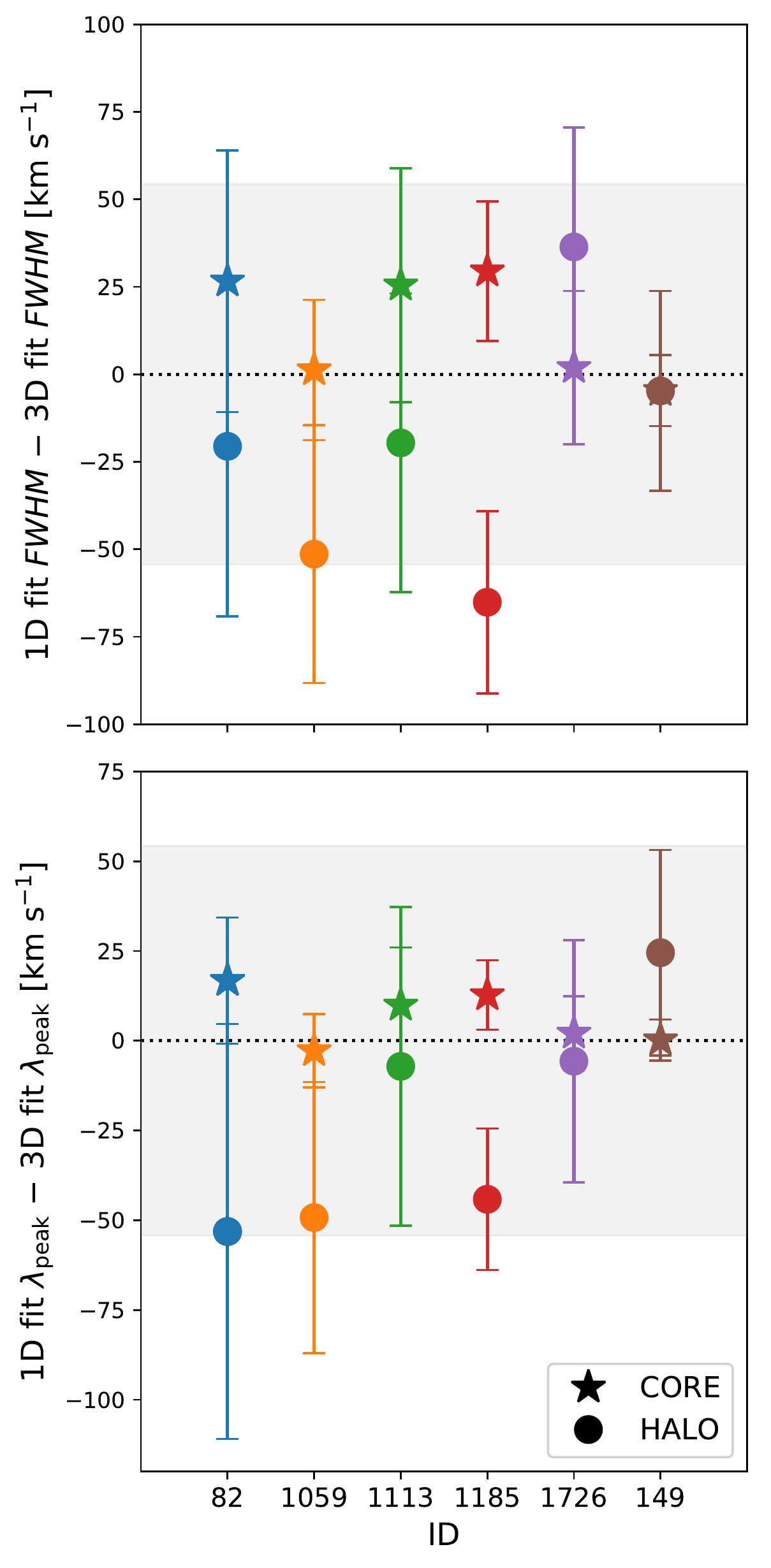}}
    \caption{Comparison between the core (stars) and halo (dots) line parameters (\textit{top}: \mbox{FWHM}, \textit{bottom}: peak position) resulting from our 3D fit method and 1D fit performed on the $\lya$ lines extracted in the central region (r < 0\farcs4) and outer regions (r > 1\arcsec) of the LAH. This exercise is performed for the six brightest LAHs of our sample (same color coding as Fig.~\ref{fig:2}). The grey shaded area indicates a difference of less than one MUSE pixel (set for z = 3).}
    \label{fig:3b}
\end{figure}

Figure~\ref{fig:3b} provides a comparison between the parameter values resulting from the 3D fit procedure and direct measurements using 1D fit. To do so, we extract a core-like (circular aperture of radius < 0\farcs4) and halo-like (annular aperture of radius > 1") line from the subcubes of the six S/N > 10 LAHs and fit those lines with an asymmetric Gaussian function (see Sect.~\ref{sec:32}). 
The peak velocity shift and \mbox{FWHM} values of the two methods are identical within the errorbars and the differences are smaller than one MUSE pixel error (grey shaded area where the error value in km s$^{-1}$ is set at $z$ = 3) for most of the galaxies. The largest differences are observed for the objects \#1059 and \#1185 and are discussed in Sect.~\ref{sec:72}). We also note that the central flux of the halo component is not taken into account for the 1D halo line measurements. This could be responsible for the resulting differences between the two methods.

We underline here that we do not expect the agreement to be perfect, considering that a simple parametric model cannot capture the complex structure and kinematics of the CGM in its entirety, but the fact that we find very good agreement in most cases is reassuring and means that deviations due to additional complexity are not dramatic and that our simple modeling does capture the overall properties of the haloes.

\section{Spectral characterization of the $\lya$ emission} 
\label{sec:5}

\begin{figure*}
\centering
   \resizebox{\hsize}{!}{\includegraphics{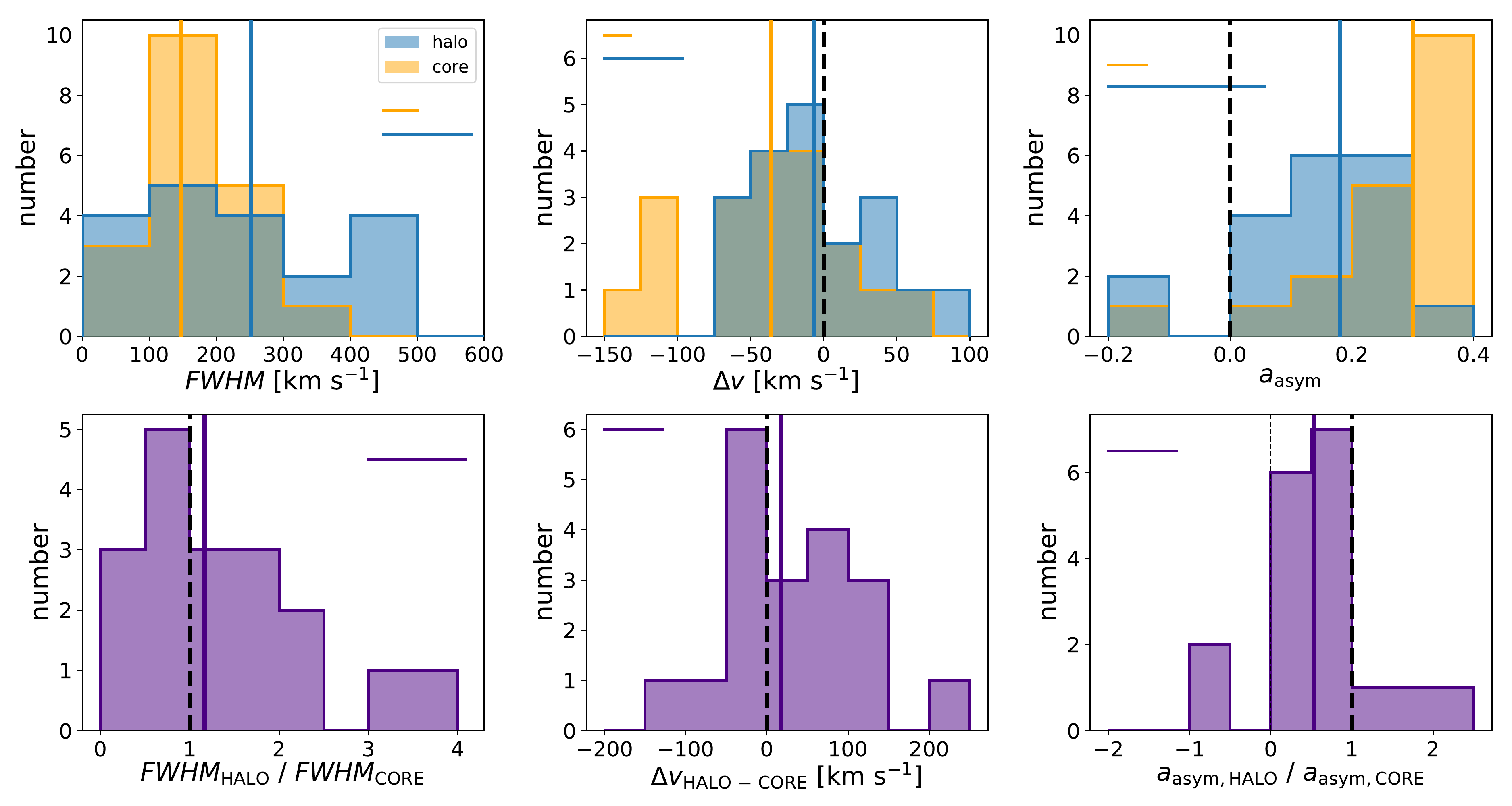}}
    \caption{Distributions of the best-fit parameters in the halo (blue) and core (orange) components (top) and their ratio or difference (bottom). The median values are indicated by the vertical solid lines. The typical errors (median values) are indicated by the horizontal solid lines in each panel.
    \textit{Left}: Distributions of the \mbox{FWHM} parameters. The median values for the core, halo and ratio are 147 km s$^{-1}$, 252 km s$^{-1}$ and 1.16, respectively. \textit{Middle}: Distributions of peak positions in the core and halo component relative to the peak of the total $\lya$ line (top) and halo/core peak separation (bottom). The median values are $-$36 km s$^{-1}$, $-$6 km s$^{-1}$ and $+$17 km s$^{-1}$, respectively. \textit{Right}: Distributions of the asymmetry parameters. The median values for the core, halo and ratio are 0.30, 0.18, and 0.5, respectively. For visibility, the object (\#7159) with halo/core \mbox{FWHM} ratio of 8.2 and $a_{\rm asym}$ ratio of $-$19.2 is not shown here.}
    \label{fig:4}
\end{figure*}

\subsection{Line parameters}
\label{sec:51}

The adopted model provides parameters reflecting the width, the peak wavelength and the asymmetry of the $\lya$ lines corrected for LSF for the two components. 
The top row of Fig.~\ref{fig:4} shows the distributions of the core (orange) and halo (blue) line \mbox{FWHM} (left), their peak velocity shift (middle) relative to the peak position of the total fitted line (purple line in Fig.~\ref{fig:3}) and their asymmetry parameters (right). The median values are indicated by the vertical solid lines. 

The \mbox{FWHM} values of the $\lya$ lines span a range from 40 to 325 km s$^{-1}$ (median value of 147 km s$^{-1}$) in the core component and from 33 to 471 km s$^{-1}$ in the halo (median value of 252 km s$^{-1}$).  
The velocity shift $\Delta v$ values between the peak of the $\lya$ line in a given component and the peak of the total $\lya$ line range from $-$142 to $+$75 km s$^{-1}$ in the core and $-$53 to $+$93 km s$^{-1}$ in the halo with a respective median value of $-$36 and $-$6 km s$^{-1}$.
The median asymmetry parameter values are 0.30 and 0.18 and vary from $-0.15$ to 0.39 and from $-$0.20 to 0.36 for the core and halo $\lya$ lines, respectively.

\subsection{Halo-Core line profiles comparison}
\label{sec:52}

Focusing on the median values of the line parameters in the core and halo taken separately (solid vertical lines in the top panels of Fig.~\ref{fig:4}), the median $\lya$ line in the halo appears broader, redder, and less asymmetric than the median $\lya$ line in the core but still asymmetric with a red tail. 
It is, however, important to note that those distributions show large dispersions (see Sect.~\ref{sec:51}).

Now, when comparing the median $\lya$ line parameters between the components of each galaxy (bottom panels), we see that those differences are less obvious. Indeed, the $\lya$ line in the halo appears only slightly broader (median value of the \mbox{FWHM} ratio of 1.16, bottom left) and slightly redder (+17 km s$^{-1}$, bottom middle) than in the core component. On the contrary, the direct core/halo comparison, with a median asymmetry parameter ratio of 0.5 (bottom right), reinforces the fact that the line in the halo is less asymmetric.
It is again important to note here that the dispersion of the distributions is high. The $\lya$ line in the halo of our galaxies can be up to about two times broader but also narrower than in the core. Some objects have a blueshifted (up to $\approx$100 km\,s$^{-1}$) or redshifted (up to 200 km\,s$^{-1}$) $\lya$ line in the halo compared to the line in the core. We also find that the $\lya$ line can be more asymmetric in the halo or even show an opposite asymmetry profile (ratio < 0) compared to the line in the core.

Those results demonstrate that the $\lya$ lines in the halo and in the core can have very different spectral profiles in terms of width, wavelength peak position and asymmetry.

\subsection{Halo-Core parameters correlation analysis}
\label{sec:53}

\begin{figure}
\centering
   \resizebox{\hsize}{!}{\includegraphics{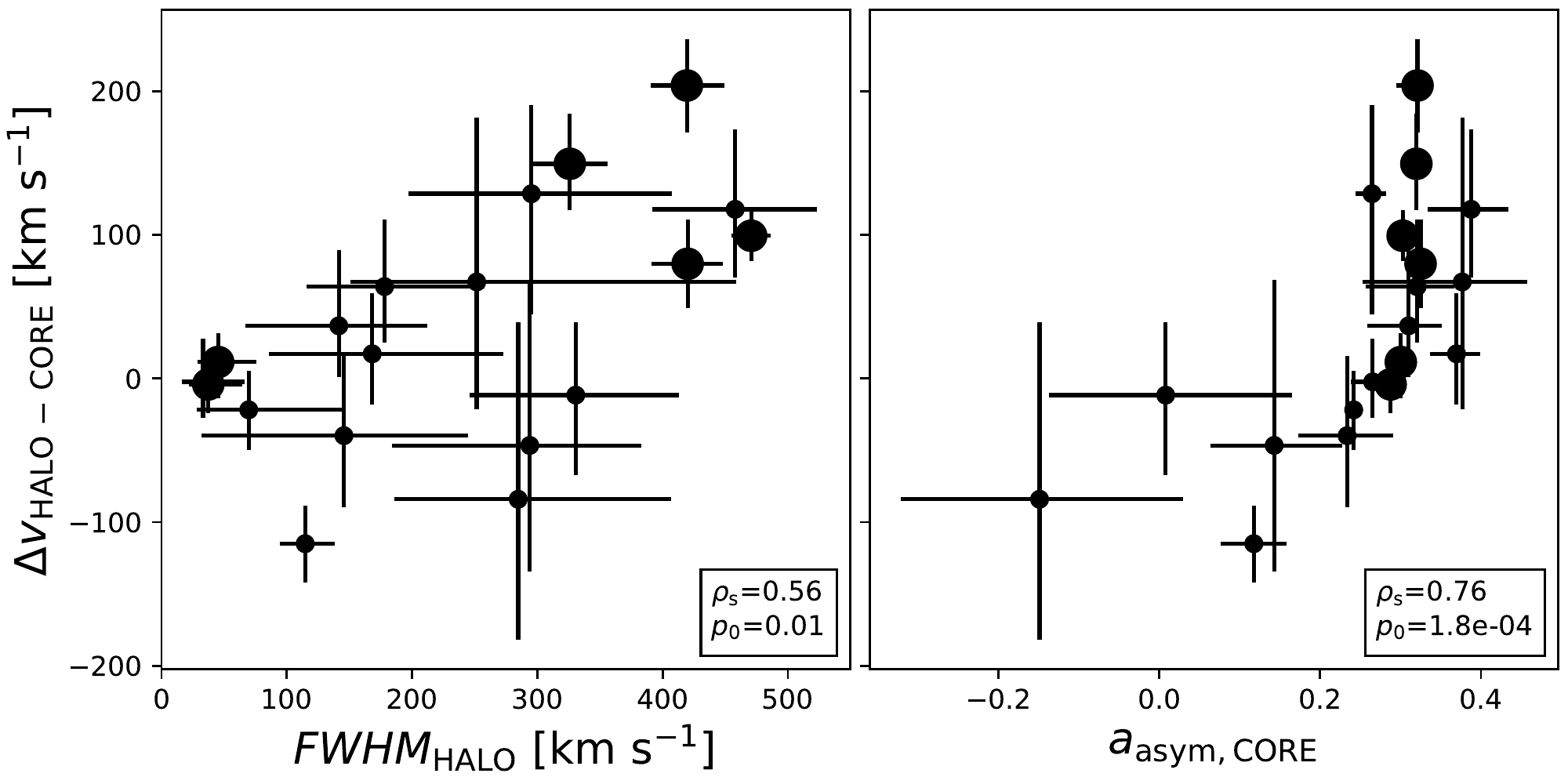}}
    \caption{Peak velocity shift of the $\lya$ line extracted in the halo relative to the peak position of the $\lya$ line extracted in the core component plotted as a function of the \mbox{FWHM} of the line in the halo (left) and the asymmetry parameter of the line in the core (right). The larger symbols indicate the six S/N$_{\rm HALO}$ > 10 objects. Spearman rank correlation coefficients $\rho_{\rm s}$ and corresponding $p_{\rm 0}$ values are shown in each panel. This figure shows the relations between the best-fit line parameters resulting from our 3D fit procedure (see Sect.~\ref{sec:413}) for which a correlation is found. The other relations are shown in Fig.~\ref{fig:ap1}.}
    \label{fig:5}
\end{figure}

We show the most significant correlations between the line parameters in Fig.~\ref{fig:5}. The relations between all best-fit parameters are displayed in Fig.~\ref{fig:ap1}.

The results of the Spearman rank correlation test ($\rho_{\rm s} = 0.56$,  $p_{\rm 0} = 0.01$, left panel) suggest that a positive correlation exists between the width and the peak velocity shift of the $\lya$ line in the halo: the wider the $\lya$ line in the halo is the more it is redshifted compared to the $\lya$ line in the core. This result is in good agreement with our spatially resolved analysis of the $\lya$ haloes (Sect.~\ref{sec:33}).

We also find that the peak velocity shift of the $\lya$ line in the halo relative to the peak position in the core correlates with the asymmetry parameter of the central $\lya$ line ($\rho_{\rm s} = 0.76$,  $p_{\rm 0} \simeq 10^{-4}$, right panel). In other words, the more asymmetric the $\lya$ line in the core is, the more redshifted the $\lya$ line in the halo is. 
This correlation is not seen in the halo component.

We do not find any other significant correlation between the line parameters (see Fig.~\ref{fig:ap1}).

\subsection{Statistical significance of the spectral differences}
\label{sec:54}

In order to answer the question of whether the $\lya$ line is significantly different in the central region and in the CGM of our galaxies, we calculate the probability $p_{\rm0}$ of the two line parameter sets (\mbox{FWHM}, $\lambda_0$, and $a_{\rm asym}$ of the core and halo components) to be identical by considering normal distributions for the parameters with dispersion equal to the statistical error on the best fit. We consider the parameters as statistically different if $p_{\rm0}$ < 10$^{-5}$. 

Out of the 19 objects of our sample, seven (37\%) show statistical evidence for a difference in their core/halo $\lya$ spectral profiles in terms of width, peak position and asymmetry. 
Considering the \mbox{FWHM}, $\lambda_0$ and $a_{\rm asym}$ parameters separately, 74\% (14 objects), 84\% (16 objects) and 79\% (15 objects) of our sample show statistically significant differences, respectively. The galaxies for which the core and halo $\lya$ lines are not statistically different are designated with open circles in Figs.~\ref{fig:6b} and \ref{fig:7}.


\section{Connecting host galaxies to their $\lya$ lines}
\label{sec:6}

In this section we investigate the relations between the spectral and spatial properties of the $\lya$ haloes (Sect.~\ref{sec:61}). Then we connect the LAH spectral characteristics to the UV content of the host galaxies (Sect.~\ref{sec:62}). The figures of this section aim to illustrate the tentative correlations found. The other relations are shown in Appendix~\ref{ap:1} (Figs.~\ref{fig:ap2} and \ref{fig:ap2b}).

\subsection{$\lya$ spatial extent and flux}
\label{sec:61}

We start by investigating the relation between the spectral (in terms of \mbox{FWHM}, asymmetry, and peak separation) and spatial (in terms of scale length and flux) characteristics of the $\lya$ emission; in each component first (Sect.~\ref{sec:611}) and then between the core and halo components (Sect.~\ref{sec:612}).

\subsubsection{Connection in each component}
\label{sec:611}

\begin{figure}
\centering
   \resizebox{\hsize}{!}{\includegraphics{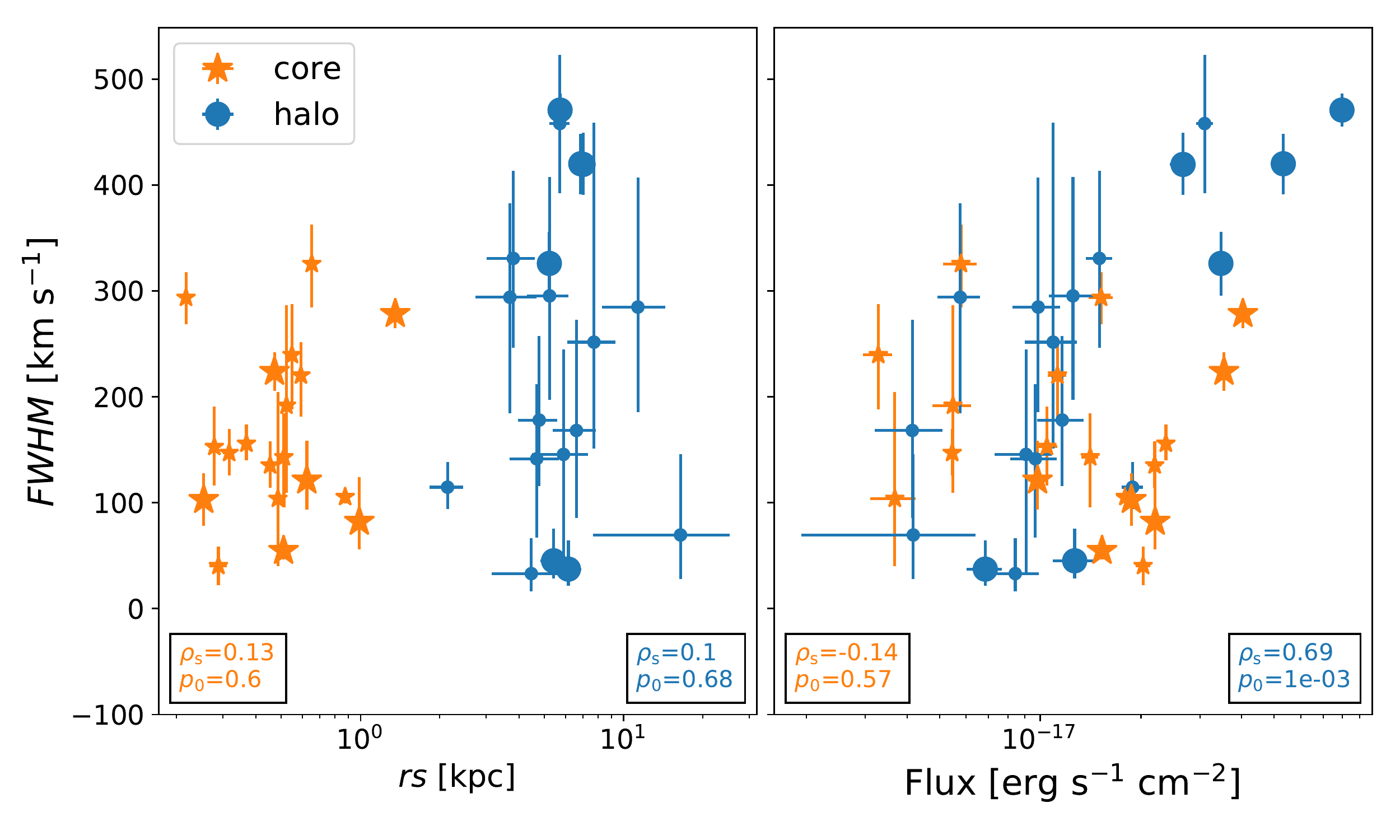}}
    \caption{Comparison between the spectral and spatial $\lya$ properties in each component taken separately. The \mbox{FWHM} of the $\lya$ line in the core (orange) and in the halo (blue) is plotted against the $\lya$ emission scale length in the left panel and against the flux in the component in the right panel. The larger symbols indicate the six S/N$_{\rm HALO}$ > 10 objects. Spearman rank correlation coefficients $\rho_{\rm s}$ and corresponding $p_{\rm 0}$ values are shown in each panel. Figure~\ref{fig:ap2} (panels a3 and a5) show the relations with the peak velocity shift in the halo and the asymmetry where no correlation is found.}
    \label{fig:6a}
\end{figure}

Figure~\ref{fig:6a} shows the relations between the $\lya$ spectral and spatial properties in terms of \mbox{FWHM} and, exponential scale length and flux, respectively. 
We find no correlation between the scale length ($rs$) and the width of the line neither in the core component nor in the halo (left panel). A lack of correlation is also observed between the scale lengths and the asymmetry parameter as well as the peak separation of the lines (see panels a3 and a5 of Fig.~\ref{fig:ap2}, respectively).
In the halo component, we find a positive correlation ($\rho_{\rm s} = 0.69$,  $p_{\rm 0} = 10^{-3}$) between the $\lya$ flux and the width of line (and therefore the peak separation as those properties appear correlated, see Sect.~\ref{sec:52} and panel c5 of Fig.\ref{fig:ap2}) which is not found in the core component (Fig.~\ref{fig:6a}, right panel).

\subsubsection{Connection between components} 
\label{sec:612}

\begin{figure}
\centering
   \resizebox{\hsize}{!}{\includegraphics{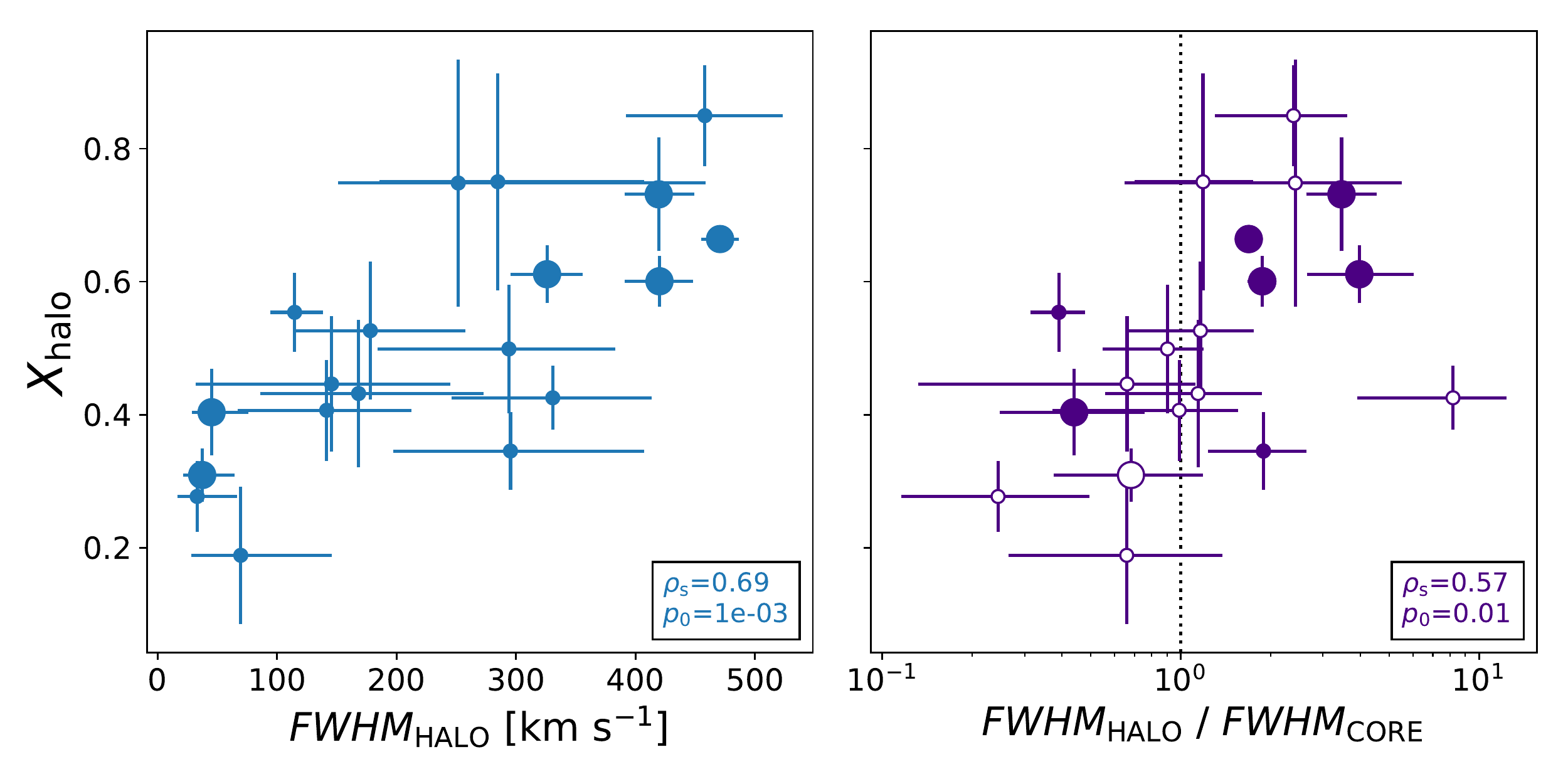}}
    \caption{Halo flux fraction plotted against the \mbox{FWHM} of the line in the halo (left) and halo/core \mbox{FWHM} ratio (right). The larger symbols indicate the six S/N$_{\rm HALO}$ > 10 objects. Spearman rank correlation coefficients $\rho_{\rm s}$ and corresponding $p_{\rm 0}$ values are shown in each panel. Open circles represent the objects for which the core/halo $\lya$ lines are not statistically different in terms of width, peak position and asymmetry parameter (see Sect.~\ref{sec:54}). 
    See also Fig.~\ref{fig:ap2} for the relations between the spatial and spectral properties of the LAHs.}
    \label{fig:6b}
\end{figure}

Next we compare the $\lya$ emission properties between the core and the halo of each galaxy. 
Figure~\ref{fig:6b} displays the halo flux fraction as a function of the \mbox{FWHM} of the line in the halo (left) and halo/core \mbox{FWHM} ratio (right). The halo flux fraction $X_{\rm halo}$ is defined as the ratio between the $\lya$ flux in the halo and the total $\lya$ flux. We find a connection between $X_{\rm halo}$ and the width of the $\lya$ line ($\rho_{\rm s} = 0.69$, $p_{\rm 0} = 10^{-3}$) as well as a suggestive correlation between $X_{\rm halo}$ and the halo/core \mbox{FWHM} ratio ($\rho_{\rm s} = 0.57$,  $p_{\rm 0} = 0.01$).

We find no link between the spatial extent of the LAHs (normalized to the UV-continuum spatial extent) and the spectral properties of the $\lya$ emission (see Fig.\ref{fig:ap2}, column b). 

In addition to the lack of $\lya$ halo size evolution with redshift already presented in L17 (see Sect.~6.3 of L17), we find no significant evolution of the $\lya$ halo spectral properties (see last column of Fig.~\ref{fig:ap2b}).

\subsection{UV properties of the host galaxy}
\label{sec:62}

\begin{figure}
\centering
   \resizebox{.7\hsize}{!}{\includegraphics{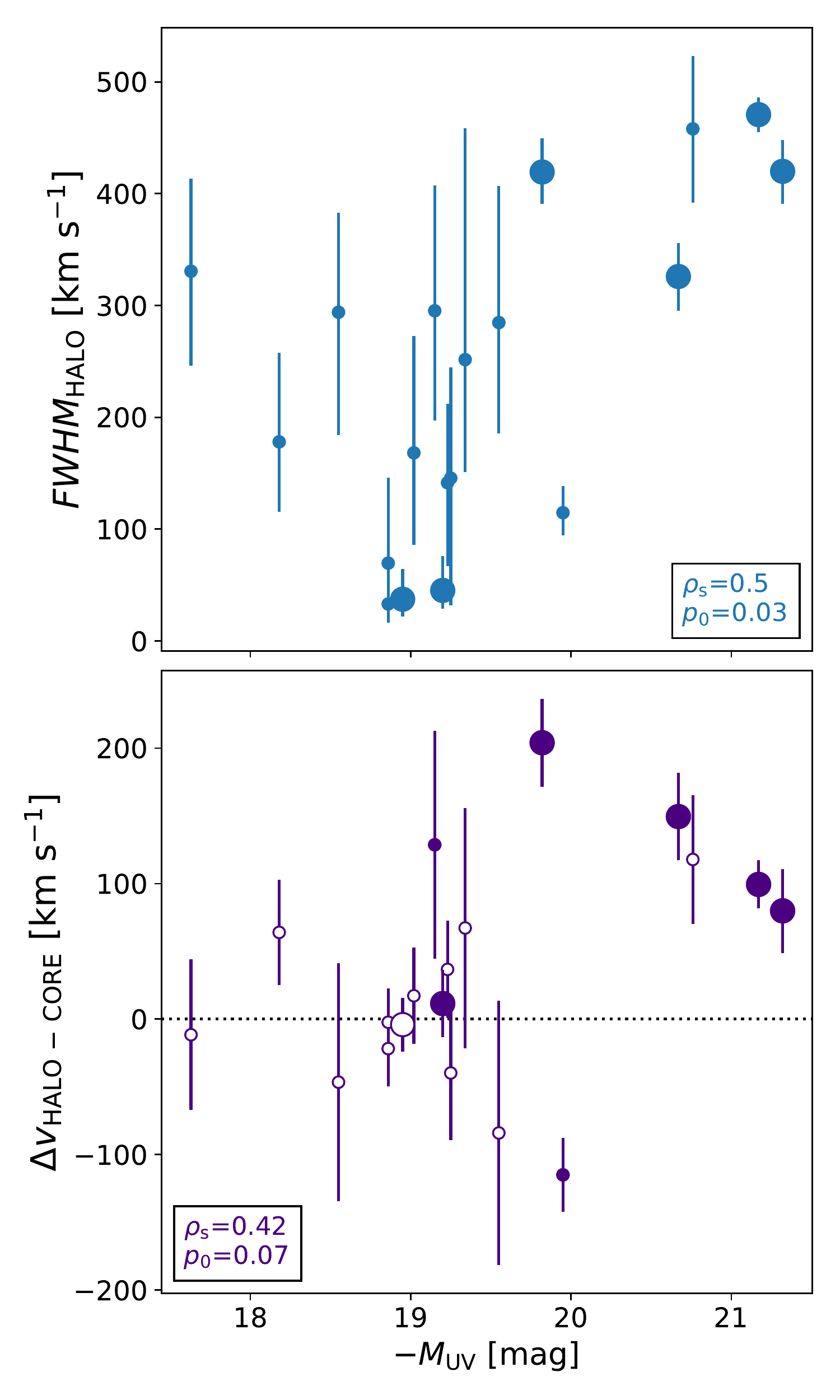}}
    \caption{$\lya$ halo spectral properties in terms of \mbox{FWHM} (top) and peak shift relative to the central line (bottom) plotted against the absolute UV magnitude of the host galaxy. The larger symbols indicate the six S/N$_{\rm HALO}$ > 10 objects. Spearman rank correlation coefficients $\rho_{\rm s}$ and corresponding $p_{\rm 0}$ values are indicated. Open circles (bottom) represent the objects for which the core/halo $\lya$ lines are not statistically different (see Sect.~\ref{sec:54}). See also Fig.~\ref{fig:ap2b} for other relations between the $\lya$ line properties and the UV properties of the host galaxy.}
    \label{fig:7}
\end{figure}

\begin{figure}
\centering
   \resizebox{\hsize}{!}{\includegraphics{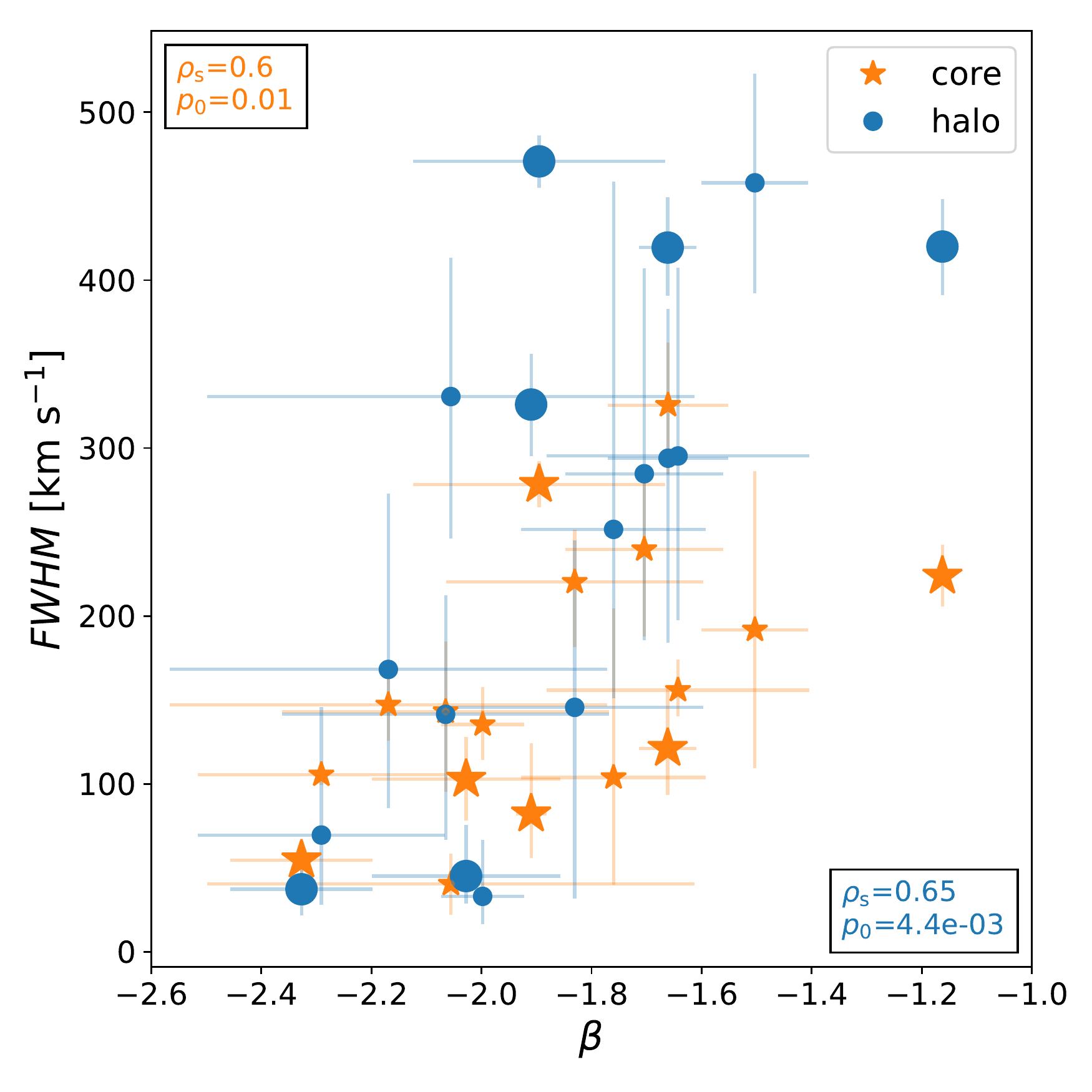}}
    \caption{FWHM of the $\lya$ line in the core (orange stars) and in the halo (blue points) plotted against the UV continuum slope of the host galaxy (calculated in H17). The larger symbols indicate the six S/N$_{\rm HALO}$ > 10 objects. The Spearman rank correlation coefficients $\rho_{\rm s}$ and corresponding $p_{\rm 0}$ values are indicated. See also Fig.~\ref{fig:ap2b} for other relations between the $\lya$ line properties and the UV properties of the host galaxy.}
    \label{fig:7b}
\end{figure}

Here we investigate the connection between the measured $\lya$ line parameters and the host galaxy properties in terms of total rest-frame $\lya$ equivalent width $EW_{\rm 0}$, absolute UV magnitude $M_{\rm UV}$, and UV continuum slope $\beta$.

While we find no strong link between the $\lya$ line parameters in the halo and either $EW_{\rm 0}$ or $M_{\rm UV}$ (first and second columns of Fig.~\ref{fig:ap2b}), there is a suggestion of a correlation between the UV magnitude of the host galaxy and the \mbox{FWHM} of the $\lya$ line extracted in the halo ($\rho_{\rm s} = 0.5$,  $p_{\rm 0} = 0.03$, top panel of Fig.~\ref{fig:7}). This is not the case for the line in the core component (see orange symbols in the second top panel of Fig.~\ref{fig:ap2b}).

Interestingly, the $\lya$ line of the haloes surrounding bright galaxies (M$_{\rm UV} < -20$ mag) are among the broadest (\mbox{FWHM}$_{\rm HALO}$ > 300 km s$^{-1}$); and they are both broader and redder than the line in the inner regions (\mbox{FWHM}$_{\rm HALO} /$ \mbox{FWHM}$_{\rm CORE}$ > 1.7 and $\Delta v_{\rm HALO - CORE}$ > 80 km s$^{-1}$, see bottom panel of Fig.~\ref{fig:7}). They also have a red tail (a$_{\rm asym, HALO}$ > 0) and are less asymmetric than the line in the galaxy core (a$_{\rm asym, HALO}$ / a$_{\rm asym, CORE}$ < 1). We do not find such trends with the equivalent width (see first column of Fig.~\ref{fig:ap2b}). 

The most significant correlation is found between the width of the line and the UV continuum slope (calculated in H17) as shown in Fig.~\ref{fig:7b}. This trend applies to the $\lya$ lines in the core (orange) and in the halo (blue) component ($\rho_{\rm s}$ $\simeq$ 0.75, $p_{\rm 0}$ $\simeq$ 5$\times 10^{-4}$). 
The $\beta$ slope is known to be steeper for galaxies with low dust content \citep{M99}, young stellar age or low metallicity. 
Our results may therefore suggest that more evolved galaxies, i.e., dustier, more metal-rich, and potentially more massive \citep{F09,Y10,O10}, show broader $\lya$ lines.

Finally, we find no correlation between the $\lya$ halo spectral profiles and the UV spatial extent of the host galaxies (i.e., $rs_{\rm core}$, see orange symbols in the first column of Fig.~\ref{fig:ap2}). The Spearman rank correlation coefficients of the two relations which are not shown in Fig.~\ref{fig:ap2} (\mbox{FWHM$_{\rm HALO}$} $-$ $rs_{\rm core}$ and $a_{\rm asym, HALO} - rs_{\rm core}$) are ($\rho_{\rm s} = 0.36$,  $p_{\rm 0} = 0.13$) and ($\rho_{\rm s} = 0.09$,  $p_{\rm 0} = 0.7$), respectively.


\section{Discussion}
\label{sec:7}

\subsection{Diversity of the $\lya$ line profiles in the halo} 
\label{sec:71}

This study allows for the first statistical analysis of the spectral signature of the $\lya$ emission in the CGM of LAEs. We find that the 
typical red-asymmetric and single-peaked $\lya$ profile is observed in the central region as well as in the outskirts of most of our galaxies. No clear hint for a blue-asymmetric profile -- that would be indicative of cooling inflows (\citealt{D06} and Sect.~\ref{sec:732}) -- is found. This result shows that either (i) if the LAHs are produced by scattering from a central source (see Sect.~\ref{sec:731}), radiative transfer effects in the CGM preserve the global spectral shape of the $\lya$ line emerging from the ISM, or (ii) if the $\lya$ photons are created in-situ in the CGM through fluorescence processes (see Sect.~\ref{sec:733}), the physical conditions in the ISM and CGM are similar.

The large dispersion of the line parameters distributions (see Fig.~\ref{fig:4}) demonstrates the large diversity of the $\lya$ halo spectral profiles. 
Together with our previous L17 study, our results show that the $\lya$ haloes around galaxies appear very diverse in terms of spatial and spectral properties.
Indeed, when comparing with the central $\lya$ spectral signatures, the $\lya$ lines in the halo are on average redder, wider and more symmetrical (Fig.~\ref{fig:4}, top).  
However, the comparison of lines in the core and halo on a galaxy by galaxy basis (see bottom panels of Fig.~\ref{fig:4}, see also Figs.~\ref{fig:5} and \ref{fig:ap1}) shows no significant correlation.
This underlines the diversity of physical properties of the CGM, and the complexity of their connection with the host galaxies.

Recently, \citeauthor{S19} (2019, hereafter S19) studied a simulated low-mass galaxy at $z \simeq$ 5 and predicted that the $\lya$ line profile does change with time but also with the viewing angle. Although their $\lya$ line profiles are double peaked, their results suggest that the large diversity observed in our sample of LAHs could be explained by the strong variability of the $\lya$ escape fraction with time and line of sight. 
\cite{Be19} reported similar results for a 10$^{10}$ M$_{\odot}$ simulated galaxy at $z = 7.2$ by looking at 48 different lines of sight. In particular, they found that the $\lya$ \mbox{FWHM} varies a lot (from 60 to 1200 km\,s$^{-1}$, see their Fig.\,3).

Interestingly, such a diversity is also appreciable at smaller scales, i.e., within the spatial extent of the LAHs (see Fig.~\ref{fig:1}). Those intrahalo variations reveal the complex structure of the cold CGM around high redshift star forming galaxies, as also pointed out by S19.

\subsection{Is the $\lya$ line really broader in the halo ?}
\label{sec:72}

The detection of small-scale variations in the $\lya$ line profile actually questions our two-component model. Indeed, for objects showing strong spatial variations in their spectral shapes, the $\lya$ line extracted from the halo component can be artificially broadened through averaging. This is probably the case for the objects \#1059 and \#1185 (see Fig.~\ref{fig:3b}, red and orange points). 
While this possibility challenges our model, the results between our resolved and parametric approaches are in relatively good agreement for most of our tested objects (see Fig.~\ref{fig:3b}). 

Additionally, the line shape is not necessarily due to radiative transfer effects, but can also include contributions from intrinsic patterns of the CGM. A CGM scale rotation is also observed in simulations (S19). This seems to be the case for object \#1059 which appears to contain a velocity gradient in the 2D map. Although the systemic redshift is unknown, we calculated the dispersion induced by the observed velocity gradient. On average 50\% of the FWHM of the $\lya$ line in the halo can be explained by the peak position variations for our six S/N > 10 objects. We thus find that a velocity field can not explain entirely the measured \mbox{FWHM} values.

The presence of close satellites can also artificially create small scale variations within the halo and therefore widen the resulting $\lya$ line in the halo. This possibility is discussed in Sect.~\ref{sec:734}.

\subsection{New inputs regarding the origin of the $\lya$ haloes}
\label{sec:73}

Considering our spectral analysis of the LAHs, we now attempt to address the question of their origin. Similarly to L17, we briefly review the possible powering mechanisms of the extended $\lya$ emission, discuss the observational expectations and try to connect them to our results.

\subsubsection{Scattering from a central source in an outflowing medium}
\label{sec:731}

In this scenario, the $\lya$ photons are produced by recombination or collision in the ISM of the host galaxy. The photons that succeed to escape the dusty ISM reach the CGM where they can be scattered by the neutral hydrogen and are potentially re-emitted towards the observer. This succession of events is a possible explanation of the observed $\lya$ haloes. \cite{K19} investigated the origin of the $\lya$ haloes using the stacking method and argued that this mechanism is dominant in M$_{\rm UV}$ > $-20$ LAEs at z$\simeq$2 (see their discussion).

The single-peaked and red-asymmetric spectral profile of most observed $\lya$ lines at $z >$ 3 can be explained by radiative transfer models as a signature of galactic outflows \citep{D06, V06}. More generally, the process of $\lya$ scattering from a central source in an outflowing wind medium has been well studied \citep{V06,DK12,Y16} and succeeds in reproducing most of the $\lya$ spectral profiles (eg. \citealt{Ha15,Gr17}). 
In a case of a static medium, we expect the $\lya$ spectral profile to be double peaked. The intervening gas in the Hubble flow located on the line of sight of galaxies absorbs the bluer photons and consequently attenuates the blue peak of $\lya$ line. The predicted transmission at z $\simeq$ 3.8, corresponding to the median value of our sample, for photons blueshifted by $-200$ km s$^{-1}$ from the resonance is more than 60\% \citep{L11}, meaning that double-peaked profiles should still be detected after IGM crossing. In other words, it seems more probable for our $\lya$ profiles to be explained by the presence of outflowing rather than static gas.
The sophisticated model of \citet{K18} manages to match the constraints on the CGM from absorption lines with outflowing gas, while producing extended $\lya$ nebulae. In order to match the results of W16, they however require that 100\% $\lya$ photons produced by the galaxies escape their ISM and scatter through the CGM. Although this fraction seems unrealistically high, their work suggests that scattering is one of the significant ingredients of extended $\lya$ nebulae.

An interesting result from our investigation is the correlation we find between the width and the peak velocity shift of the $\lya$ line which seems to hold for both the $\lya$ emission coming from the ISM and CGM according to our resolved maps and parametric fitting method (see top panels of Fig.~\ref{fig:2} and Fig.~\ref{fig:5}, left panel). This correlation is also found when considering integrated spectrum in models (V18) where photons are emitted from a central source and propagate in an expanding geometry, although the slopes differ. 
The spectral shapes that we report for the halo component are new constraints that could be derived from models such as that of \citet{K18}, and which may help discriminating among possible scenarios.

Moreover, if the $\lya$ photons are all produced inside galaxies and because the scattering events have a smoothing effect, the $\lya$ emission is expected to be rather homogeneous around the galaxy, as illustrated in S19 (their Fig.~9). This is however complicated to compare to our results here as the MUSE PSF also has a smoothing effect.
Looking at the spectral properties of our galaxies, the $\lya$ resolved maps (Fig.~\ref{fig:1}) reveal that some galaxies show significant spectral variations within their CGM. This observation can imply variations in the \hi\ column density, covering factor or ionization state of outflowing gas which, as suggested in numerical simulations, presents a complex structure (e.g., \citealt{M06,P19}, S19).

\subsubsection{Gravitational cooling}
\label{sec:732}

In this scenario, the $\lya$ haloes are powered by the gravitational energy of the primordial cold hydrogen gas located in the IGM and falling into the dark matter (DM) potential of the host galaxy along filaments. The mechanism at play is the so-called "cooling radiation" \citep{H00,F01,Fur05} for which the hydrogen atoms in the CGM are excited, and $\lya$ photons emitted, through collisional processes. 

Theoretical and numerical studies \citep{D06,T16} predict the spectral shape of the $\lya$ line in a cooling configuration to be blueshifted relative to the systemic redshift with a blue tail with respect to the peak of the line. 

\begin{figure}
\centering
   \resizebox{\hsize}{!}{\includegraphics{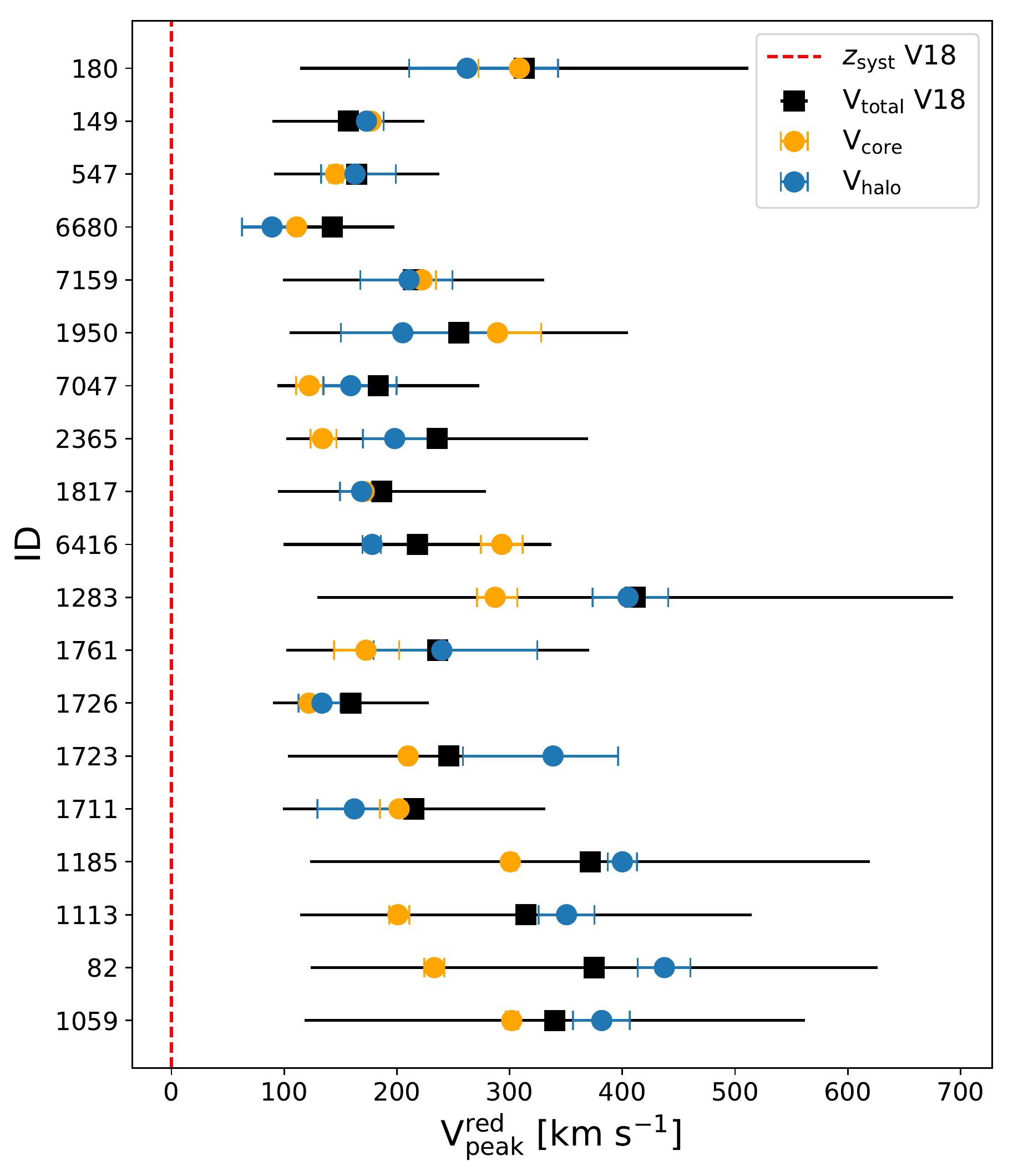}}
    \caption{$\lya$ peak velocity shift relative to the systemic redshift estimated using the V18 relation for our S/N > 6 sample. The red dashed line indicates the systemic redshift position (i.e., zero shift). The peak velocity shift of the total $\lya$ line relative to the V18 systemic redshift is indicated by the black squared symbols. The best-fit peak position of the lines in the halo and core are indicated in blue and orange, respectively.}
    \label{fig:8}
\end{figure}

Yet, this prediction is impossible to verify with our data because the MUSE spectra of our galaxies do not show any other lines enabling to measure of the systemic redshift.
However, because the vast majority of our total $\lya$ lines are single-peaked and asymmetric with a red tail, we make the assumption that they correspond to the red peak of the $\lya$ line and use the V18 empirical relation to estimate the systemic redshift of our galaxies. Our results show that, in some cases, the $\lya$ line in the core and/or halo components can be bluer ($\Delta v$ < 0, top middle panel of Fig.~\ref{fig:4} and e.g., \#82, \#1113 or \#1711 in Fig.~\ref{fig:3}, orange or blue lines) than the peak of the total $\lya$ line (purple line in Fig.~\ref{fig:3}) which is the line used to establish the V18 relation. 
This velocity shift reaches $-$142 km s$^{-1}$ in the core and $-$53 km s$^{-1}$ in the halo component. In the following, we want to figure out if those lines are bluer than the systemic redshift.
The V18 relation predicts the peak of the total $\lya$ line to be redshifted from the systemic redshift by values ranging from $+$143 km s$^{-1}$ ($\pm$ 55 km s$^{-1}$) to $+$411 km s$^{-1}$ ($\pm$ 282 km s$^{-1}$) for our sample. We used the \mbox{FWHM} measured from the best-fit total $\lya$ line not corrected for the MUSE LSF because the V18 relation is not corrected for instrumental effects.
The peak velocity shifts of the 19 objects with S/N > 6 calculated using the V18 relation are shown in Fig.~\ref{fig:8} (black squared symbols).
Considering such empirical systemic redshift values, Fig.~\ref{fig:8} shows that the $\lya$ core or halo lines (orange and blue symbols) do not appear blueshifted (even when considering the 1$\sigma$ errors) relative to the systemic redshift (indicated by the red dashed line).
Interestingly, some LAEs seem to show asymmetric lines with a blue tail mainly in the halo (e.g., \#149, \#1723 or \#7159, Fig.~\ref{fig:3} or Fig.~\ref{fig:4} top right). If those lines are actually blueshifted, it would be interpreted by the models as a hint of cooling gas flow. 

In conclusion, our test for this scenario is clearly limited by the fact that we do not know the systemic redshift of our galaxies.
As mentioned in our sample definition (Sect.~\ref{sec:22}), 16 double-peaked LAEs have been excluded for this study. The blue bumps of those objects may correspond to this expected cooling signal predicted by models. Such objects will be studied in future work. However those spectral features are not common in the MUSE UDF field compared to the single-peaked LAEs at $z$ > 3 and therefore, this scenario does not appear to be dominant. Finally, we note that interestingly some objects (e.g., \#1950, \#6416) have a $\lya$ line which is more blueshifted in the halo than in the core. A better spectral resolution is needed here to determine if this could actually be non-resolved blue dominant double-peaked lines and therefore a hint of inflowing cold gas. 

\subsubsection{Fluorescence}
\label{sec:733}

This scenario assumes that there is enough Lyman continuum (LyC) radiation escaping the host galaxy to photoionize the surrounding hydrogen gas in the CGM. Additional ionizing radiation could also be generated by external sources like a nearby quasar or the cosmic UV background \citep{Fur05,C05,K10}. This mechanism is usually invoked for large $\lya$ blobs around quasars (e.g., \citealt{Bo16}) or very compact galaxies \citep{M17}. There is no type-1 AGN in our sample (see Sect.~5 of H17 for discussion), and we thus assume that the ionizing sources are the young stars residing in the \hii\ regions of the host galaxies. 

In order to produce extended nebulae, this scenario requires that the CGM contains a volume-filling component of high-density neutral gas (generally referred to as clumps). The high density allows these clumps to recombine fast enough so as to convert ionizing radiation into $\lya$ photons without being instantly photo-evaporated. In such a scenario, we expect relatively little scattering effects, because the covering fraction of clumps needs to be low enough for distant clumps to be illuminated by ionizing radiation and create the observed extended $\lya$ flux. In that case, the line width and shape will be indicative of the velocity distribution of the clumps. The fact that we find broad and red lines in the haloes in general would then again suggests strong outflows with broad velocity distributions, peaking at high velocities.
As in the previous section, we note that we have excluded from our sample haloes with double peaked lines. In the present scenario, these may have been signs of gas clumps falling in. This is perhaps what we see in object \#149 (see Fig.~\ref{fig:1}).

A powerful diagnostic to test this scenario is to look at the H$\alpha$ emission. Indeed, if we find it to be extended, it would mean that $\lya$ photons are emitted from a photo-ionized medium and thus through a fluorescence process (\citealt{MR17}, their Fig.~6). \cite{So17} did compare the physical scales of their stacked H$\alpha$ and $\lya$ images constructed from LAEs at z$\simeq$2 and found the H$\alpha$ distribution to be two times smaller. Those results suggest that this scenario is not (the only one) at play. We note however that their LAEs sample has a slightly brighter luminosity range (42.5 < log($L_{\rm \lya}$) [erg s$^{-1}$] < 43.5) and that the sources with the highest $\lya$ luminosities are dominated by X-ray detected AGN.

\subsubsection{Satellites}
\label{sec:734}

Numerical simulations show that galaxies are rarely alone but often surrounded by a number of satellite galaxies located inside their virial radius (e.g., \citealt{SU10, L15, S19}). In this scenario, the $\lya$ halo would actually consist of the $\lya$ radiation emitted by satellites \citep{Mo16}. 

The analytical study of \cite{MR17} has shown that satellite LAEs located at large radii (> 20 kpc) from the central galaxy can contribute to the powering of the $\lya$ haloes at those radii. Because our resolved analysis (Sect.~\ref{sec:3}) focuses on a more central region of the halo (< 20 kpc), we thus do not expect a contribution from satellites for our haloes.  
However, recent resolved $\lya$ halo studies using lensing clusters \citep{P16, C19} reveal the presence of one satellite associated to the central galaxy. This satellite is responsible for variations in the $\lya$ spectral profile (see their resolved maps) and is located at $\approx$10 kpc from the central galaxy. One can then wonder whether the spatial variations detected within our LAHs indicate the presence of satellites. After careful visual inspection using deep HST imaging\footnote{We look at HST band including $\lya$ emission. We also check the MUSE and photometric redshift of the surrounding sources}, we do not find any strong evidence of such objects except maybe for the \#1185 LAE. Indeed, looking at its peak position map (Fig.~\ref{fig:1}), the blue feature on the right of the source center seems to correspond to one of the galaxies visible in the HST image. One of those HST galaxies has a close redshift ($z_{\rm BPZ}$ = 4.46$\pm$0.25 and $z_{\rm EAZY}$ = 4.36$\pm$0.10) corresponding to a velocity shift of $\approx$ $-$100 km s$^{-1}$ which is consistent with the observed peak velocity shift.
Moreover, it is interesting to note that the presence of a satellite is suspected in the vicinity of the second lensed LAE analyzed in C19. This suspicion is based on the fact that strong spectral variations are detected in a localized region at $\approx$8 kpc from the central galaxy. The authors found no HST counterpart for this prospective satellite highlighting that such satellites could be very faint in UV continuum. This possibility would agree with \cite{MR17} who predict the UV absolute magnitude of satellites to be > $-17$. 

In fact, with the same data as used in the present paper, I17 report a large number of LAEs with no HST counterpart in the catalog of \cite{R15}. \cite{M18} further investigated these objects and showed that their typical UV continuum magnitude is about $-15$. These findings remind us that despite the exquisite depth of Hubble data in our field, we can probably not see the UV counterparts of small satellite galaxies that may contribute significantly to the LAHs. However, the typical $\lya$ flux of these objects is $\approx$10$^{-18}$ $\flcgs$ and so one would require large numbers of satellites ($\gtrsim$10) to account for the full Lya luminosity of the haloes. We also note that the velocity distribution of satellite galaxies is unlikely to produce a red asymmetric line, and thus argue that the contribution of faint companions to the LAHs is probably not dominant.


\section{Summary and conclusions}
\label{sec:8}

From our integral field spectroscopic observations of the LAEs detected in the MUSE UDF, we analyzed the spectral properties of 19 LAHs detected around individual star-forming galaxies at redshift $z$ > 3. Those objects were selected to have a LAH with a good S/N (> 6). Two methods were employed: we constructed spatially resolved maps of the $\lya$ spectral properties for the six brightest LAHs (Sect.~\ref{sec:3}) and performed a 3D two-component fit including the 13 fainter ones (Sect.~\ref{sec:4}). This study allowed us to push further the analysis of the LAHs and better characterize the properties of the CGM. Our main results are summarized as follows: 
\begin{enumerate}
    \item 
    We detect small-scale variations in the spectral shape of the $\lya$ emission between the halo and the core regions of the galaxies but also within the halo itself (Figs.~\ref{fig:1} and \ref{fig:3}). The observed $\lya$ profiles span a large range of parameters in terms of width and peak velocity shift, and are different for each object (Sect.~\ref{sec:33} and \ref{sec:51}). This result emphasizes the complexity and the diversity of configurations of the cold CGM (Sect.~\ref{sec:71}).
    \item 
    We find that the $\lya$ lines in the center of the galaxy and in the CGM are statistically different (in terms of peak velocity shift, width and asymmetry) for $\approx$40\% of our tested objects (Sect.~\ref{sec:54}). 
    \item
    On average, the $\lya$ line in the haloes are broader, redder and less asymmetric than the $\lya$ line in the cores (see Fig~\ref{fig:4}, upper panels). However, the dispersion of the line parameters are very broad. In particular, we find objects that show an opposite trend: narrower and bluer line in the halo.
    \item 
    We find a relation between the width of the LAH line and its peak velocity shift (Fig.~\ref{fig:5}, left panel). This result holds globally as well as locally (see Fig.~\ref{fig:2}, top row) on individual objects. A correlation between the asymmetry of the central $\lya$ line and its velocity separation with the line in the halo is also found (Fig.~\ref{fig:5}, right panel).
    \item 
    We investigated the relation between the spectral and spatial properties of the LAHs (Sect.~\ref{sec:61}). While we found no correlation with the scale lengths (see Fig.~\ref{fig:6a}, left panel), our results show that the galaxies with high $\lya$ flux fraction in the halo have a broader $\lya$ line in the halo than in the core (Fig.~\ref{fig:6b}). 
    \item
    We do not observe any significant evolution of the $\lya$ line parameters between the redshift range of 3 to 6. 
    \item
    Interestingly we found that the UV bright ($M_{\rm UV}$ < $-$20 mag) galaxies of our sample have the broadest $\lya$ lines in the halo (Fig.~\ref{fig:7}, top panel). Additionally, their $\lya$ lines are broader, redder and less asymmetric in the halo. Finally, the most significant correlation we found is between the $\beta$ slope of the host galaxy and the \mbox{FWHM} of the emerging $\lya$ line in the core and in halo (Fig.~\ref{fig:7b}). This result suggests that the dustier galaxies have broader $\lya$ lines.
    \item 
    With the new information we obtained from our spectral analysis of the LAHs, we attempted to answer the question of their origin. 
    The generally broad and red line shapes found in the halo component suggests either scattering through an outflowing medium, fluorescent emission from outflowing cold clumps of gas, or a mix of both. Gravitational cooling, because of the blue line it is expected to produce, appears unlikely to explain our observations. An important contribution of satellites could be possible with no counterpart in very deep UV images, but we argue that it is unlikely to be dominant because of the asymmetry of the observed lines. 
\end{enumerate}

Our MUSE data allowed us to perform the first statistical analysis of the spatially resolved spectral properties of the $\lya$ distribution around individual and distant star forming galaxies. This study emphasizes the large amount of information encoded in the $\lya$ spectral shape as well as the difficulty to decode it. Detailed comparisons of observations with realistic simulations as well as larger and deeper LAE samples are needed to have a clearer idea of the mechanisms regulating the gas exchanges between the galaxies and their CGM at $z>3$.


\begin{acknowledgements}
F.L., R.B. and S.C. acknowledge support from the ERC advanced grant 339659-MUSICOS. F.L., T.G., H.K. and A.V. acknowledge support from the ERC starting grant ERC-757258-TRIPLE. A.C. and J.R. acknowledge support from the ERC starting grant 336736-CALENDS. J.B. acknowledges support by FCT/MCTES through national funds (PID-DAC) by grant UID/FIS/04434/2019 and through Investigador FCT Contract No.IF/01654/2014/CP1215/CT0003. T.H. was supported by Leading Initiative for Excellent Young Researchers, MEXT, Japan.
\end{acknowledgements}


\bibliographystyle{aa} 
\bibliography{biblio}

\begin{appendix}

\onecolumn

\section{Looking for correlations}
\label{ap:1}

\begin{figure*}[h!]
\centering
   \resizebox{\hsize}{!}{\includegraphics{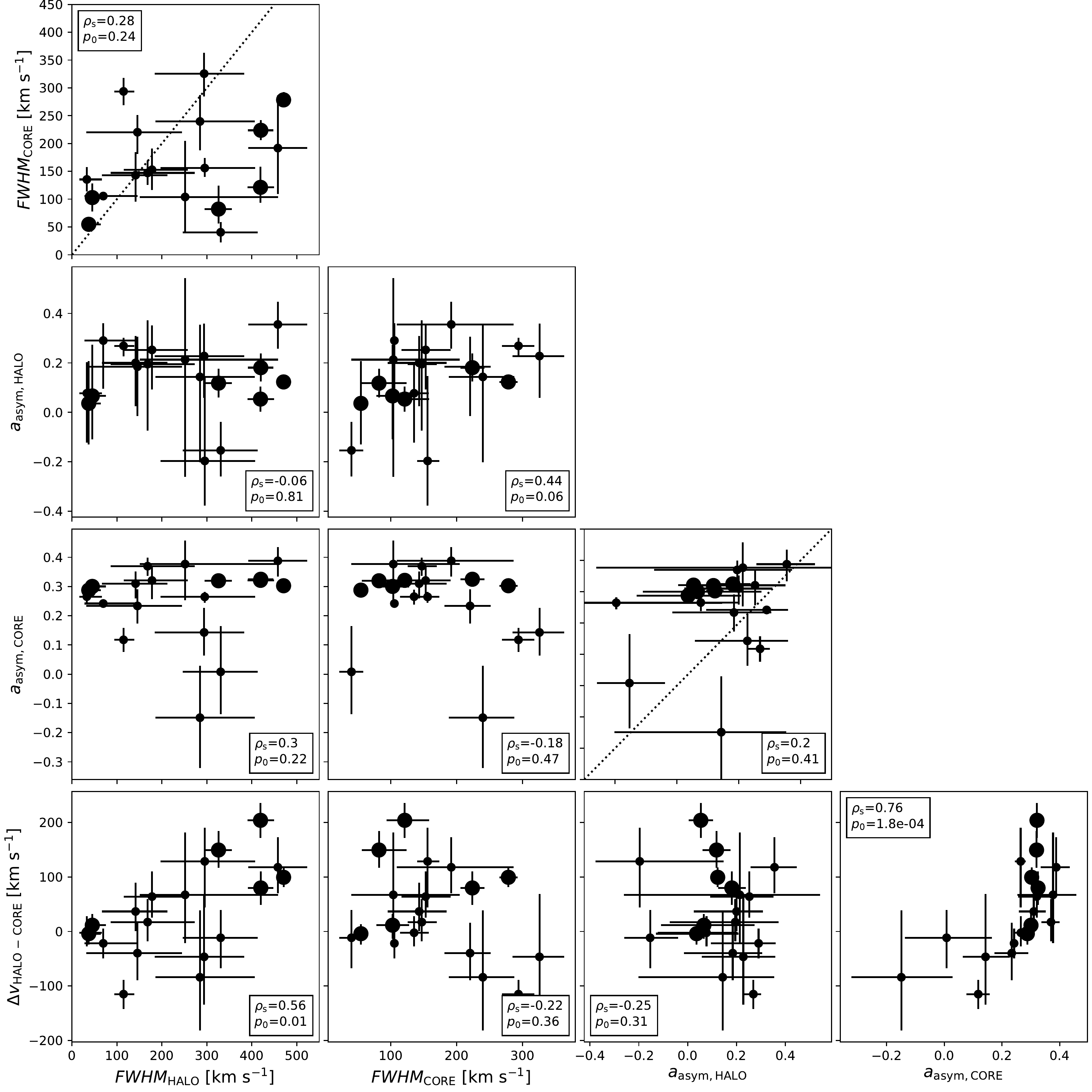}}
    \caption{Relations between the line parameters resulting from our 3D fit procedure (see Sect.~\ref{sec:413}). The larger symbols indicate the six S/N$_{\rm HALO}$ > 10 objects. Spearman rank correlation coefficients $\rho_{\rm s}$ and corresponding $p_{\rm 0}$ values are shown in each panel. The dotted lines represent the one to one relation. The first and last panels of the bottom row are also displayed in Fig.~\ref{fig:5}.}
    \label{fig:ap1}
\end{figure*}

\begin{figure*}[h!]
\centering
   \resizebox{\hsize}{!}{\includegraphics{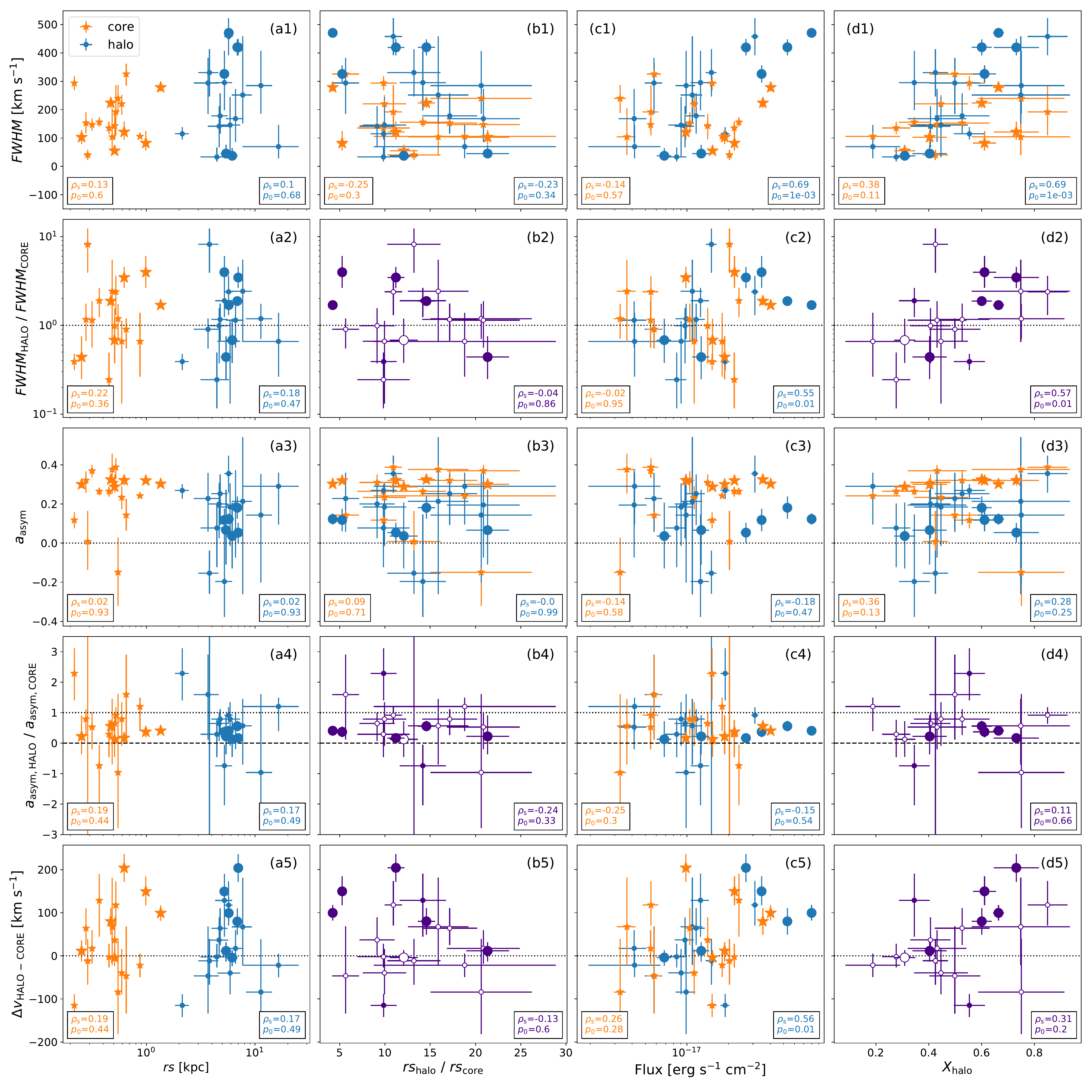}}
    \caption{$\lya$ emission spectral properties (from top to bottom: \mbox{FWHM}, halo/core \mbox{FWHM} ratio, asymmetry parameter a$_{\rm asym}$, halo/core a$_{\rm asym}$ ratio and peak separation) plotted against its spatial properties (from left to right: scale lengths, scale length ratio, flux, halo flux fraction). The larger symbols indicate the six S/N$_{\rm HALO}$ > 10 objects. Spearman rank correlation coefficients $\rho_{\rm s}$ and corresponding $p_{\rm 0}$ values are shown in each panel. 
    In order to ease the reading, the rows are numbered from 1 to 5 and the columns designated by letters from a to d.
    Open circles dots represent the objects for which the core/halo $\lya$ lines are not statistically different in terms of width, peak position and asymmetry parameter (see Sect.~\ref{sec:54}).}
    \label{fig:ap2}
\end{figure*}

\begin{figure*}[h!]
\centering
   \resizebox{\hsize}{!}{\includegraphics{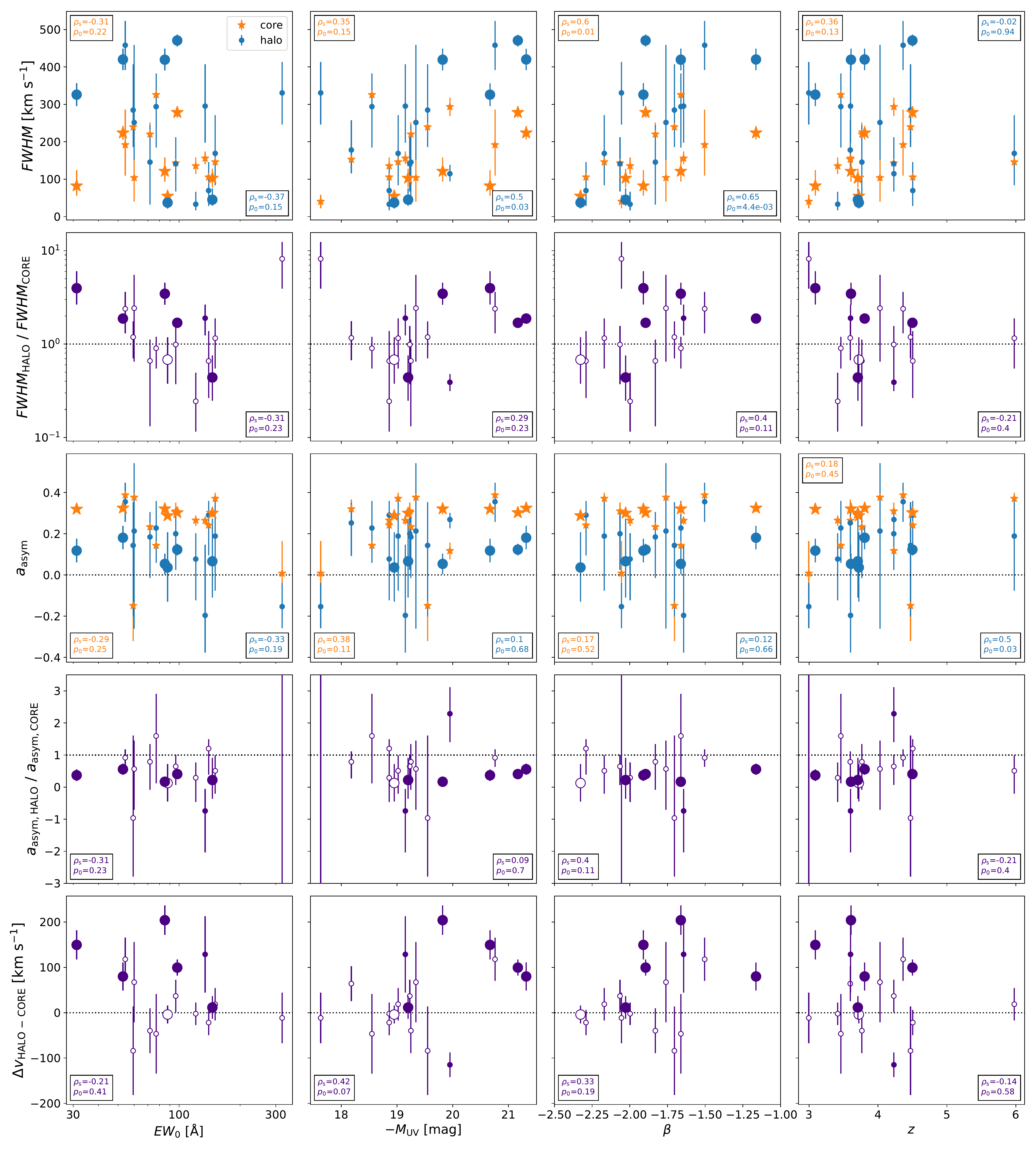}}
    \caption{$\lya$ emission spectral properties (same as Fig.~\ref{fig:ap2}) plotted against, from left to right : the $\lya$ $EW_{\rm 0}$, the absolute UV magnitude, the UV continuum slope and the redshift of the host galaxy. The larger symbols indicate the six S/N$_{\rm HALO}$~>~10 objects. Spearman rank correlation coefficients $\rho_{\rm s}$ and corresponding $p_{\rm 0}$ values are indicated. Open circles represent the objects for which the core/halo $\lya$ lines are not statistically different (see Sect.~\ref{sec:54}).}
    \label{fig:ap2b}
\end{figure*}

The following figures show the different comparisons that we did in order to investigate the correlations between : 
\begin{itemize}
    \item the spectral properties of the LAHs (Fig.~\ref{fig:ap1})
    \item the spectral and spatial properties of the LAHs (Fig.~\ref{fig:ap2})
    \item the spectral properties of the LAHs and the UV properties and redshift of the host galaxy (Fig.~\ref{fig:ap2b}).
\end{itemize}

\newpage
~
\newpage
~
\newpage

\section{Best-fit parameters not corrected for LSF effects}
\label{ap:2}

The LSF of the MUSE data has been characterized in \cite{B17}. Table~\ref{tab:1b} reports the best-fit parameters values not corrected by the MUSE LSF.  
On average the LSF has a broadening effect of $\approx$100 km s$^{-1}$ and a redshift effect of $\approx$20 km s$^{-1}$ (median values) in the core component (no effect on the median value are observed for the lines extracted in the halo component). 

\begin{table*}[h!]
\caption{Fitting results from our 3D two-component analysis on 19 LAEs not corrected for LSF effects.}
\centering
\def\arraystretch{1.5}
\begin{tabular}{cccccccc}
\hline
\hline
$\rm ID$ & FWHM$_{\rm CORE}$ & FWHM$_{\rm HALO}$ & $a_{\rm asym,CORE}$ & $a_{\rm asym,HALO}$ & $\lambda \rm peak_{\rm CORE}$ & $\lambda \rm peak_{\rm HALO}$ & $\Delta \rm v_{\rm HALO-CORE}$ \\
& [km s$^{-1}$] & [km s$^{-1}$] &  &  & [\AA\ ] & [\AA\ ] & [km s$^{-1}$] \\
\hline
82 & 263$_{-15}^{+16}$ & 464$_{-26}^{+27}$ & 0.27$_{-0.02}^{+0.02}$ & 0.06$_{-0.04}^{+0.04}$ & 5600.1$_{-0.1}^{+0.1}$ & 5603.5$_{-0.4}^{+0.4}$ & 181$_{-28}^{+28}$ \\
149 & 210$_{-5}^{+5}$ & 211$_{-21}^{+23}$ & 0.13$_{-0.01}^{+0.01}$ & -0.09$_{-0.07}^{+0.07}$ & 5738.1$_{-0.0}^{+0.1}$ & 5737.8$_{-0.3}^{+0.3}$ & -15$_{-18}^{+17}$ \\
180 & 386$_{-29}^{+30}$ & 388$_{-60}^{+65}$ & 0.11$_{-0.06}^{+0.06}$ & 0.15$_{-0.12}^{+0.11}$ & 5421.3$_{-0.5}^{+0.5}$ & 5420.8$_{-0.9}^{+1.2}$ & -30$_{-78}^{+95}$ \\
547 & 224$_{-12}^{+14}$ & 241$_{-72}^{+107}$ & 0.29$_{-0.03}^{+0.03}$ & 0.11$_{-0.20}^{+0.20}$ & 8479.9$_{-0.2}^{+0.2}$ & 8480.2$_{-0.8}^{+0.8}$ & 10$_{-35}^{+34}$ \\
1059 & 330$_{-8}^{+9}$ & 471$_{-25}^{+25}$ & 0.25$_{-0.01}^{+0.01}$ & 0.17$_{-0.04}^{+0.04}$ & 5841.0$_{-0.1}^{+0.1}$ & 5842.0$_{-0.4}^{+0.4}$ & 55$_{-23}^{+24}$ \\
1113 & 301$_{-15}^{+16}$ & 381$_{-25}^{+26}$ & 0.25$_{-0.02}^{+0.02}$ & 0.14$_{-0.04}^{+0.04}$ & 4971.0$_{-0.1}^{+0.1}$ & 4972.8$_{-0.3}^{+0.3}$ & 104$_{-24}^{+24}$ \\
1185 & 340$_{-8}^{+9}$ & 507$_{-14}^{+15}$ & 0.26$_{-0.01}^{+0.01}$ & 0.11$_{-0.02}^{+0.02}$ & 6683.8$_{-0.1}^{+0.1}$ & 6685.6$_{-0.2}^{+0.2}$ & 80$_{-14}^{+15}$ \\
1283 & 324$_{-60}^{+72}$ & 496$_{-49}^{+53}$ & 0.34$_{-0.07}^{+0.07}$ & 0.28$_{-0.08}^{+0.08}$ & 6519.0$_{-0.4}^{+0.5}$ & 6521.6$_{-0.6}^{+0.6}$ & 117$_{-44}^{+51}$ \\
1711 & 292$_{-18}^{+19}$ & 288$_{-53}^{+58}$ & 0.18$_{-0.04}^{+0.04}$ & 0.14$_{-0.12}^{+0.11}$ & 5792.4$_{-0.2}^{+0.2}$ & 5791.7$_{-0.7}^{+0.8}$ & -41$_{-47}^{+57}$ \\
1723 & 272$_{-8}^{+10}$ & 429$_{-103}^{+137}$ & 0.17$_{-0.02}^{+0.02}$ & -0.19$_{-0.14}^{+0.18}$ & 5592.2$_{-0.1}^{+0.1}$ & 5593.5$_{-0.7}^{+0.7}$ & 70$_{-42}^{+44}$ \\
1726 & 220$_{-11}^{+12}$ & 214$_{-22}^{+24}$ & 0.19$_{-0.03}^{+0.03}$ & 0.19$_{-0.11}^{+0.08}$ & 5720.8$_{-0.1}^{+0.1}$ & 5720.5$_{-0.3}^{+0.3}$ & -19$_{-20}^{+20}$ \\
1761 & 246$_{-49}^{+67}$ & 332$_{-66}^{+141}$ & 0.32$_{-0.15}^{+0.14}$ & -0.02$_{-0.28}^{+0.33}$ & 6110.3$_{-0.5}^{+0.6}$ & 6112.0$_{-1.2}^{+1.1}$ & 81$_{-82}^{+80}$ \\
1817 & 261$_{-11}^{+11}$ & 212$_{-38}^{+46}$ & 0.16$_{-0.02}^{+0.02}$ & 0.05$_{-0.11}^{+0.11}$ & 5366.3$_{-0.1}^{+0.1}$ & 5365.9$_{-0.4}^{+0.4}$ & -24$_{-27}^{+26}$ \\
1950 & 290$_{-35}^{+39}$ & 327$_{-65}^{+86}$ & -0.12$_{-0.12}^{+0.12}$ & 0.15$_{-0.20}^{+0.17}$ & 6650.6$_{-0.7}^{+0.6}$ & 6648.9$_{-1.0}^{+1.3}$ & -80$_{-77}^{+86}$ \\
2365 & 287$_{-21}^{+23}$ & 290$_{-51}^{+76}$ & 0.15$_{-0.05}^{+0.06}$ & 0.27$_{-0.12}^{+0.11}$ & 5587.4$_{-0.2}^{+0.3}$ & 5587.9$_{-0.5}^{+0.6}$ & 25$_{-40}^{+44}$ \\
6416 & 339$_{-24}^{+23}$ & 225$_{-19}^{+24}$ & 0.10$_{-0.04}^{+0.04}$ & 0.21$_{-0.04}^{+0.04}$ & 6358.9$_{-0.4}^{+0.4}$ & 6357.0$_{-0.2}^{+0.2}$ & -91$_{-26}^{+27}$ \\
6680 & 193$_{-3}^{+3}$ & 269$_{-142}^{+93}$ & 0.12$_{-0.01}^{+0.01}$ & 0.20$_{-0.48}^{+0.18}$ & 6689.6$_{-0.0}^{+0.0}$ & 6688.9$_{-1.0}^{+1.0}$ & -32$_{-46}^{+45}$ \\
7047 & 258$_{-16}^{+16}$ & 186$_{-31}^{+40}$ & 0.19$_{-0.03}^{+0.03}$ & 0.22$_{-0.08}^{+0.08}$ & 6355.4$_{-0.2}^{+0.2}$ & 6355.5$_{-0.4}^{+0.5}$ & 4$_{-26}^{+31}$ \\
7159 & 221$_{-9}^{+13}$ & 414$_{-77}^{+64}$ & 0.02$_{-0.04}^{+0.04}$ & -0.08$_{-0.10}^{+0.09}$ & 4856.3$_{-0.1}^{+0.1}$ & 4855.6$_{-0.7}^{+0.6}$ & -46$_{-47}^{+45}$ \\
\end{tabular}
\tablefoot{ID: source identifier in the MUSE UDF catalog by I17. FWHM$_{\rm CORE}$: rest-frame full width at half maximum of the $\lya$ line extracted in the core component in km s$^{-1}$. FWHM$_{\rm HALO}$: rest-frame full width at half maximum of the $\lya$ line extracted in the halo component in km s$^{-1}$. $a_{\rm asym,CORE}$: asymmetry parameter of the $\lya$ line in the core. $a_{\rm asym,HALO}$: asymmetry parameter of the $\lya$ line in the halo. $\lambda \rm peak_{\rm CORE}$: peak wavelength position of the $\lya$ line in the core in \AA. $\lambda \rm peak_{\rm HALO}$: peak wavelength position of the $\lya$ line in the halo in \AA. $\Delta \rm v_{\rm HALO-CORE}$: halo/core peak separation in km s$^{-1}$.}

\label{tab:1b}
\end{table*}

\end{appendix}

\end{document}